%% file: Machine_learning_based_plasticity.tex
\documentclass[a4paper]{paper_cw}

\usepackage{helvet}
\usepackage{amsmath}
\usepackage{amssymb}
\usepackage{mathrsfs}
\usepackage{graphics}
\usepackage{natbib}
\usepackage{color}
\usepackage{fancyhdr} 
\usepackage{colortbl} 
\usepackage{subfigure}
\usepackage[dvips]{graphicx} 
\usepackage{marvosym}
\usepackage{natbib}
\usepackage{import}
\include{definitions}

\usepackage{graphicx}

\begin{document}

\title{A machine learning based plasticity model using proper orthogonal decomposition}

\author{Dengpeng Huang \and Jan Niklas Fuhg \and Christian Wei{\ss}enfels \and Peter Wriggers
}

\institute{D. Huang (\Letter), J. N. Fuhg, C. Wei{\ss}enfels, P. Wriggers  \at Institute of Continuum Mechanics, Leibniz University of Hannover, Appelstr. 11, 30167 Hannover, Germany
\\\email{ huang@ikm.uni-hannover.de}}

\maketitle
\thispagestyle{empty}

\abstract{Data-driven material models have many advantages over classical numerical approaches, such as the direct utilization of experimental data and the possibility to improve performance of predictions when additional data is available. One approach to develop a data-driven material model is to use machine learning tools. These can be trained offline to fit an observed material behaviour and then be applied in online applications. However, learning and predicting history dependent material models, such as plasticity, is still challenging. In this work, a machine learning based material modelling framework is proposed for both elasticity and plasticity. The machine learning based hyperelasticity model is developed with the Feed forward Neural Network (FNN) directly whereas the machine learning based plasticity model is developed by using of a novel method called Proper Orthogonal Decomposition Feed forward Neural Network (PODFNN). In order to account for the loading history, the accumulated absolute strain is proposed to be the history variable of the plasticity model. Additionally, the strain-stress sequence data for plasticity is collected from different loading-unloading paths based on the concept of sequence for plasticity. By means of the POD, the multi-dimensional stress sequence is decoupled leading to independent one dimensional coefficient sequences. In this case, the neural network with multiple output is replaced by multiple independent neural networks each possessing a one-dimensional output, which leads to less training time and better training performance. To apply the machine learning based material model in finite element analysis, the tangent matrix is derived by the automatic symbolic differentiation tool AceGen. The effectiveness and generalization of the presented models are investigated by a series of numerical examples using both 2D and 3D finite element analysis.}


\section{Introduction}
\label{S:1}

With the development of data mining technology, machine learning algorithms, high performance computing and robust numerical methods, data-driven computational modelling play an important role in not only accurate but also fast predictions of complex industrial processes. In particular, accurate material models are key parts in structure analysis. In the past years, tremendous effort has been made in developing material models, see e.g. the review of models by \cite{cao2017models} for metal forming processes. However, the proposed models show limitations in generalization or accuracy in some cases when the model is applied to engineering problems.

As a data-driven approach, the Machine Learning (ML) based material modelling provides an alternative tool to narrow the gap between experimental data and material models. By use of the ML technology, such as artificial neural networks, see e.g. \cite{hassoun1995fundamentals}, or Gaussian Processes, see e.g. \cite{rasmussen2003gaussian}, constitutive equations can be approximated by using experimental data without postulation of a specific constitutive model. An advantage of machine learning based material models is that they can iteratively be improved if more experimental data are available, which yields more flexible and sustainable material descriptions. For a review of machine learning in computational mechanics, see \cite{OISHI2017327} and references therein.

In order to replace the classical constitutive model in computational mechanics by data-driven modelling, multiple approaches have been proposed in the literature. The model-free data-driven computing paradigm proposed by \cite{kirchdoerfer2016data}, \cite{doi:10.1002/nme.5716}, \cite{EGGERSMANN201981} and \cite{stainier2019model}, conducts the computing directly from experimental material data under the constraints of conservation laws, which bypasses the empirical material modelling step. This approach works without constitutive model and seeks to find the closest possible state from a prespecified material data set. A manifold learning approach is proposed by \cite{Ibanez2017}, \cite{ibanez2018manifold} and \cite{ibanez2019hybrid}, where the so-called constitutive manifold is constructed from collected data. A self-consistent clustering approach has been developed to predict the behaviour of heterogeneous materials under inelastic deformation, see \cite{liu2016self} and \cite{shakoor2019data}. \cite{tang2019map123} proposed a mapping approach, where one-dimensional data are mapped into three-dimensions for nonlinear elastic material modelling without the construction of an analytic mathematical function for the material equation. Since the performance of the data-driven computing is highly determined by quality and completeness of the available data, data completion and data uncertainties have been investigated, see \cite{ayensa2018new} and \cite{AYENSAJIMENEZ2019120}.

In addition to the data-driven approaches mentioned above, the artificial neural network as a machine learning approach has been applied to approximate the constitutive model based on data as well, see \cite{ghaboussi1998new}, \cite{hashash2004numerical}, and \cite{lefik2003artificial}. In order to fit a constitutive material equation, the neural network is trained offline using experimental data collected from different loading paths. Afterwards, the network based model is applied online for testing and applications. A nested adaptive neural network has been applied in \cite{ghaboussi1998autoprogressive}, \cite{ghaboussi1998new} for modelling the constitutive behaviour of geomaterials. In \cite{hashash2004numerical}, a feed forward neural network based constitutive model is implemented in finite element analysis to capture the nonlinear material behaviour, where the consistent material tangent matrix is derived and evaluated. Artificial neural networks are also applied as incremental non-linear constitutive models in \cite{lefik2003artificial} for finite element applications. Furthermore, this approach has been applied to predict the stress-strain curves and texture evolution of polycrystalline metals by \cite{ALI2019}. Instead of the offline training, the neural network based constitutive model can be trained online by auto-progressive algorithms as well, see \cite{pabisek2008self} and \cite{ghaboussi1998autoprogressive}. Lastly, artificial neural networks have been applied to the heterogeneous material modelling, such as \cite{le2015computational}, \cite{lu2019data}, \cite{li2019predicting}, \cite{liu2019deep} and \cite{yang2019derivation}.

The data-driven model free approach conducts calculations directly from the data, which bypasses the model on one hand but highly relies on the quality and completeness of the data on the other hand. The machine learning approaches mentioned above apply the previous strain and stress as history variables, which introduces extra errors for the elastic stage of inelastic deformation, and thus affects the capabilities to capture the load history in real applications. Additionally, the derivation of the tangent matrix for the neural network based model is complex when changing the network architecture. Thus, there are many issues present in machine learning based material modelling approaches, such as the data collection strategy, the selection of history variables and the applications in finite element analysis.

The objective of this work is to develop a machine learning based hyperelastic and plasticity models for finite element applications as well as a corresponding data collection strategy. To simplify the data collection process from experiments, only strain components act as input data and only stress components represent output data. Instead of using the previous strain and stress as history variables in plasticity, in this work, the accumulated absolute total strain is applied as history variable to distinguish different loading paths. This variable can be computed from preexisting input data without additional effort as e.g. different experiments. Due to its history dependence, the training data for plasticity will be sequential data sets obtained under different loading-unloading paths. Since the isotropic plasticity can be formulated in the principle space, the training sequence data is collected only from tension and compression tests, which simplifies the data collection. A novel method called Proper Orthogonal Decomposition Feedforward Neural Network (PODFNN) is proposed in combination with the introduced history variable for predicting the stress sequences in case of plasticity. By means of the Proper Orthogonal Decomposition (POD), the stress sequence is transformed into multiple independent coefficient sequences, where the stress at any time step can be recovered by a linear combination of the coefficients and the basis.

The presented approach decomposes the strain-stress relationship into multiple independent neural networks with only one output, which significantly decreases the complexity of the model. In order to apply the machine learning based model in finite element analysis, the tangent matrix has to be computed. It is derived by the symbolic differentiation tool AceGen, see {\cite{korelc2016automation}. The effectiveness and generalization of the machine learning based plasticity model is validated in 2D and 3D using several applications.
	
This paper is structured as follows: In Section 2, the machine learning based material modelling framework is presented. Then the Feed forward Neural network (FNN) is applied to learn the hyper-elastic material law in Section 3. In Section 4, the data collection strategy for plasticity is proposed. Based on the training data, the machine learning based plasticity model is developed and validated in finite element analysis in Section 5, which is followed by the conclusions in Section 6.
	
\section{Data-driven material modelling framework}
To develop a data-driven material model by means of machine learning technology, three steps are necessary: data collection, machine learning and validation, see Fig. \ref{flowchart}. As a fundamental ingredient for data-driven models, the data, representing the material behaviour, have to be collected firstly. According to the specific problem, the training data can be collected from experiments and simulations. 

In this work, strain-stress data are employed as the input and output of the data-driven model. For the plasticity model, the strain-stress data will be collected for specific loading paths and stored as sequences. Depending on the problem, the training data usually have to be preprocessed utilizing data scaling, data decomposition and data arrangement.
	
The second step is to fit the constitutive equation related to the data by means of the ML technology. The Artificial Neural Network (ANN) as a machine learning technology will be employed in this work. The hyperparameters of the neural network based model have to be selected according to the data and the accompanying accuracy requirements. Once the model is trained, the describing parameters will be used and stored for the material description of the developed model.
	
The final step is to validate the accurate reproduction of the ML based material model. To do so, the ML based material model is compared with a standard material model within several finite element applications. By deriving the tangent matrix and residual vector, the ML based model can be incorporated into a FEM code. The performance of the developed model will be evaluated by benchmark tests. If the accuracy of the material model can not meet the necessary requirements, the model hyperparameters will be optimized or supplemental data will be collected. Therefore, the machine learning based framework is an open system and the accuracy of the developed model can be improved iteratively during its application.\\
	
\begin{figure}[!htb]
 \centering
 \def\svgwidth{0.8\columnwidth}
 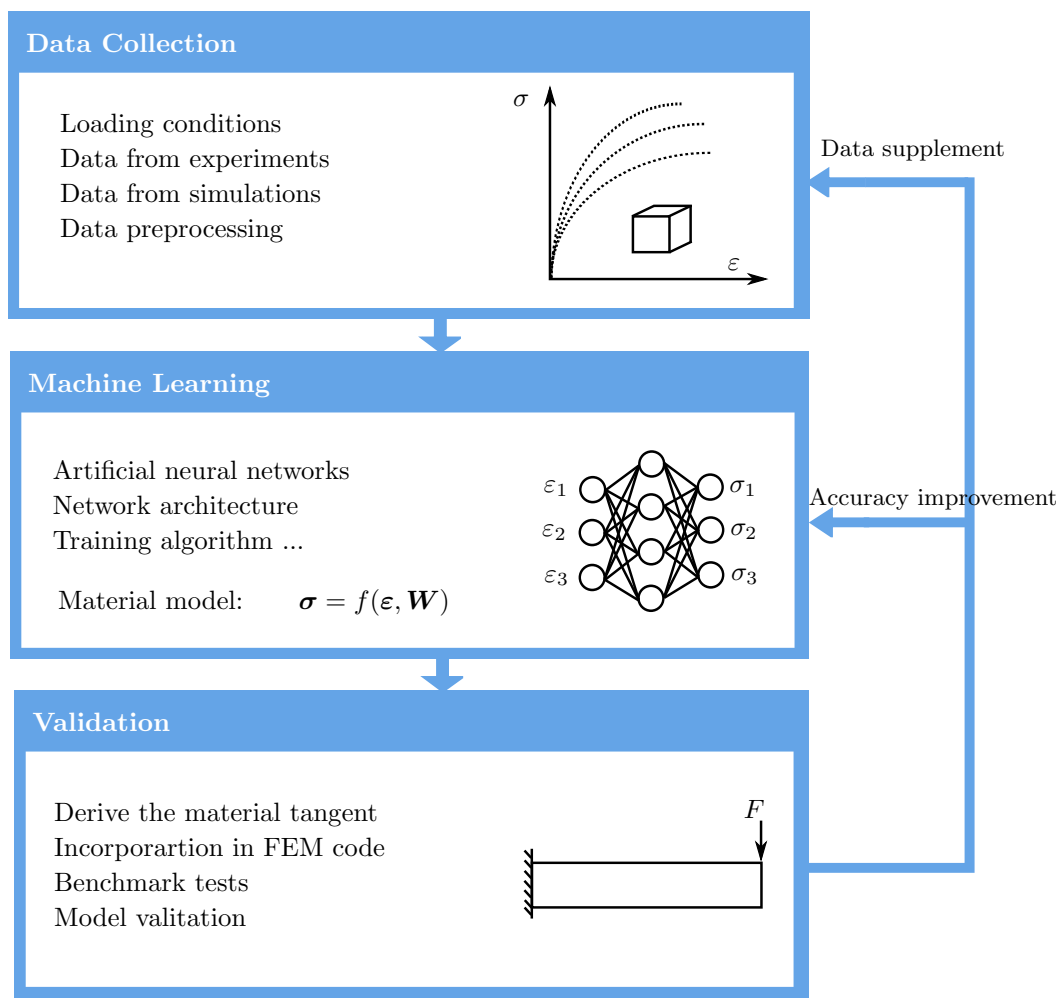
 \caption{The data-driven material modelling framework.}
 \label{flowchart}
\end{figure}
	
\section{Machine Learning (ML) based hyperelasticity}
Before the plasticity model is developed in detail, a ML based hyper-elasticity model is presuited by utilizing feed forward neural networks (FNNs) in this section.
	
\subsection{Feed forward neural network}
Feed forward neural network is a fundamental machine learning technology. A deep FNN is composed of several connected layers of artificial neurons and biases, where the data is fed into an input layer and then flows through some hidden layers. The output is finally predicted at an output layer, as shown in Fig. \ref{fnntrain}. The neurons from different layers are fully connected through the weights $w$. In the prediction phase, the data flows in one way from the input layer to the output layer. In the training phase, the global error defined by the mean-squared differences between the target value and the FNN output will be back-propagated through the hidden layers. This step is performed in order to update the weights, where the objective is to minimize the global error.\\
	
\begin{figure}[!htb]
 \centering
 \def\svgwidth{0.85\columnwidth}
 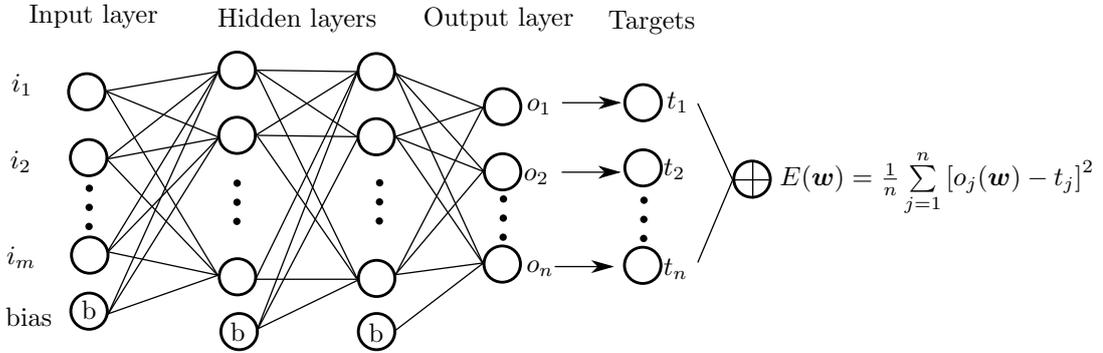
 \caption{The feed forward neural network.}
 \label{fnntrain}
\end{figure}
	
At each neuron, an activation function is attached, see Fig. \ref{neuron}. The output of each neuron is computed by multiplying the outputs from the previous layer with the corresponding weights. For the neuron $j$ in the layer $k$, the data of the previous layer $k-1$ is summed up and then altered by an activation function. The output of the neuron $j$ in layer $k$ is computed as
	\begin{align}
	o_j^k=f_s\left( \sum_{i=1}^{N} w_{ij} o_i^{k-1}+b^{k-1}_i \right),
	\end{align}
	where $N$ is the number of neurons in the previous layer $k-1$, $w_{ij}$ is the weight connecting neurons $i$ and $j$, $o_i^{k-1}$ is the output of the neuron $i$ in layer $k-1$ whereas $b_{i}^{k-1}$ is its bias. A common choice for the activation function is the sigmoid function
	\begin{align}
	f_s(x)=\frac{2}{1+e^{-2x}}-1.
	\end{align}

	\begin{figure}[!htb]
		\centering
		\def\svgwidth{0.5\columnwidth}
		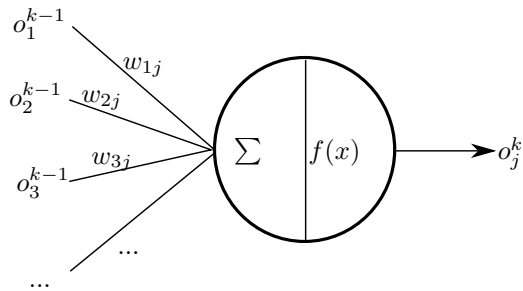
		\caption{The artificial neuron.}
		\label{neuron}
	\end{figure}
	
	The specific architecture of the FNN, such as the number of layers and the number of neurons in each layer, has to be determined according to the complexity of the data set.
	\subsection{Neural network training}
	
	In the training phase, the weighs of neural network will be initialized firstly, see \cite{nguyen1990improving}, which is followed by the weights updating using a training algorithm such that the global error is minimized. The global error, also named as loss function or network performance, is defined according to the difference between the network prediction and the target data as shown in Fig. \ref{fnntrain}. The mean squared error is used to measure the loss
	\begin{align}
	E(\boldsymbol{w})=\frac{1}{N} \sum_{i=1}^{N} \left[ o_i(\boldsymbol w)-t_i \right]^2=\frac{1}{N} \sum_{i=1}^{N}  e_i,
	\end{align}
	where $N$ is the number of outputs, $o_i$ is the $i$-th output, $\boldsymbol w$ is the vector that contains the weights of neural network, and $t_i$ is the $i$-th target value. Training a feed forward neural network is an optimization problem, where the global error is treated as the objective function. To minimize the global error, the Levenberg-Marquardt algorithm is applied to update the weighs, see \cite{hagan5menhaj},
	\begin{align}
	\boldsymbol w^{n+1} =\boldsymbol w^{n}- (\boldsymbol J^T \boldsymbol J +\mu \boldsymbol{I})^{-1} \boldsymbol J^T \cdot \boldsymbol e,
	\end{align}
	in which $\boldsymbol w^{n+1}$ is the weight vector in iteration $n+1$, $\mu$ is a parameter to adaptively control the speed of convergence, and $\boldsymbol J$ is the Jacobian matrix that contains the derivatives of network errors with respect to the network weights
	\begin{align}
	J_{ij} = \frac{\partial e_i}{\partial w_j^{n}}.
	\end{align}
	In the training process, many iterations are required to update the weights until the stopping criteria is fulfilled, where one iteration is also known as one epoch.
	
	\subsection{Data collection for the ML based hyperelasticity}
	To approximate hyperelastic behaviour by the FNN for finite element applications, the first task is to determine the input and output variables for the neural network. Since the loading and the unloading curve coincide for the elastic deformation, as shown in Fig. \ref{series1d}, the relationship between the strain space and stress space can be seen as a one-to-one mapping. Hence, the strain-stress mapping can be approximated by the FNN without considering the loading history.\\ 
	
	\begin{figure}[!htb]
		\centering
		\def\svgwidth{0.4\columnwidth}
		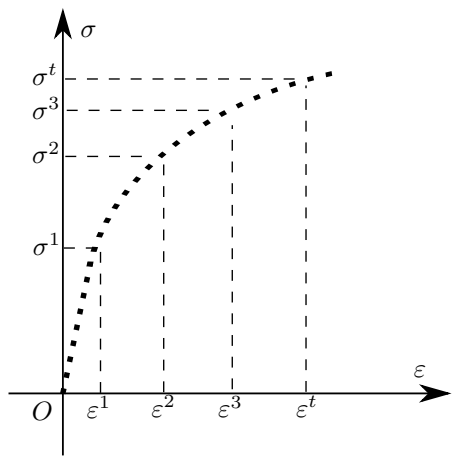
		\caption{The loading and unloading curve for hyperelasticity.}
		\label{series1d}
	\end{figure}
	
	Instead of using experimental data, the training data are collected here from an analytical model, which allows us to test the performance of the ML based model by comparing its simulation results with that by an analytical model. As an example of hyperelasticity, the non-linear neo-Hookean model is applied as the target model to learn
	\begin{align}
	\boldsymbol \sigma = \frac{1}{2}\frac{\lambda}{J}(J^2 - 1) \boldsymbol{I} + \frac{\mu}{J}(\boldsymbol{b} - \boldsymbol{I}),
	\label{neohook}
	\end{align}
	where the Cauchy stress $\boldsymbol \sigma$ and the left Cauchy Green tensor $\boldsymbol{b}$ are symmetric tensors. For the 2D problem, the inputs of the model can be chosen as the strain components $(J,b_{11}, b_{22}, b_{12})$, whereas the outputs are chosen as the stress components $(\sigma_{11}, \sigma_{22}, \sigma_{12})$. According to the number of input and the output, the architecture of the FNN can be determined as $4-n-3$ for instance, where one hidden layer with $n$ neurons is applied for this hyperelastic law. The input data is generated by taking equally spaced points within the given range of strain space. The stresses as output data can be computed from the neo-Hookean model in equation (\ref{neohook}) accordingly.
	
	Before training the FNN, the generated data is scaled to the range  $(-1,1)$ such that training is accelerated. Then the neural network is trained until the stopping criteria is reached. After training, the weights $\boldsymbol{w}$ and bias $\boldsymbol{b_s}$ will be saved as the model parameters. The ML based hyper-elasticity model is thus expressed as
	\begin{align}
	\boldsymbol\sigma^{NN}&=FNN(\boldsymbol b, J, \boldsymbol{w}, \boldsymbol b_s),
	\end{align}
	where $\boldsymbol\sigma^{NN}$ is the predicted Cauchy stress by the FNN.
	\subsection{The residual and tangent}
	The ML based model can be used in the same way as the classical constitutive model in the finite element analysis. The residual for the static problem is given by
	\begin{align}
	\boldsymbol{R} (\boldsymbol{u})=\boldsymbol{f}-\int_\Omega \boldsymbol{B}^T \boldsymbol{\sigma}^{NN} d\Omega, \\
	\boldsymbol{f}=\int_\Omega N \rho \hat{\boldsymbol{b}} dv- \int_{\partial\Omega} N \hat{\boldsymbol{t}} da,
	\end{align}
	in which $\boldsymbol{B}$ is the gradient of shape functions $N$, $\rho$ is the density, $\boldsymbol{\sigma}^{NN}$ is the stress computed from the machine learning based model, $\hat{\boldsymbol{b}}$ and $\hat{\boldsymbol{t}}$ are the body force and the surface traction respectively. Due to the non-linearity, the Newton Rapson iterative solution scheme is applied. The tangent matrix is computed by taking the derivative of residual in terms of displacement
	\begin{align}
	\boldsymbol{K}_T = \frac{\partial \boldsymbol{R}(\boldsymbol{u})}{\partial \boldsymbol u}.
	\end{align}
	
	The derivation of the tangent matrix for the neural network based model requires the computation of derivatives by the chain rule, which will be complex if the number of neuron is very large. In this work, the automatic differentiation tool AceGen, see \cite{korelc2016automation}, based on the symbolic computing in Mathematica is applied, by which the tangent matrix and residual vector can be derived automatically.
	
	\subsection{Testing the ML based hyperelasticity model in FEM}
	The material parameters for the neo-Hookean model used in the training data collection are set as $E=700N/mm^2,\nu=0.499$. An FNN with architecture of 4-10-3 is applied, with 4 neurons in the input layer, 10 neurons in the hidden layer and 3 neurons in the output layer. The Levenberg-Marquardt algorithm \cite{hagan5menhaj} is applied as the training optimizer. After 14082 training iterations, the mean squared error decreased to $0.0326$, which costs training time of $6h40m55s$.
	
	The first example is the uniaxial compression test of a plate in 2D. As shown in Fig. \ref{bargeo}, the pressure is imposed on the top surface of the plate, the bottom of the plate is fixed in vertical direction. The distributed load is given as $ q_0=-20MPa$.
	\begin{figure}[!htb]
		\centering
		\def\svgwidth{0.3\columnwidth}
		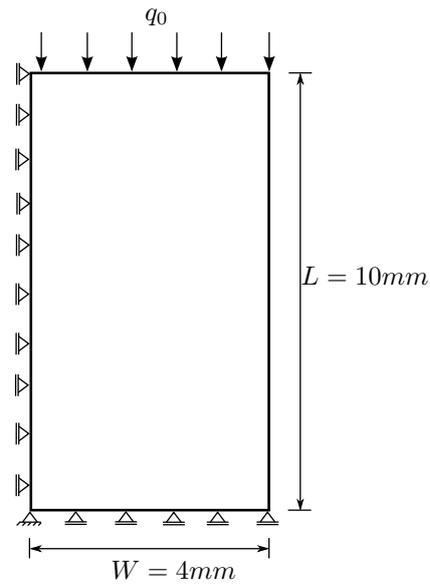
		\caption{Compression of the plate.}
		\label{bargeo}
	\end{figure}
	
	The final deformation of the plate computed with the ML based model is compared with the outcome of using the neo-Hooken model in equation (\ref{neohook}). It can be see from Fig. \ref{barneofnn} that the displacements in vertical direction are very close. The computation time with analytical hyperelastic model is $8.14s$, whereas the computation time with the ML based model is $9.75s$ on the same computer.
	
	\begin{figure}[!htb]
		\centering
		\subfigure[Neo-Hookean model]{\label{test1dplas}\includegraphics[width=60mm]{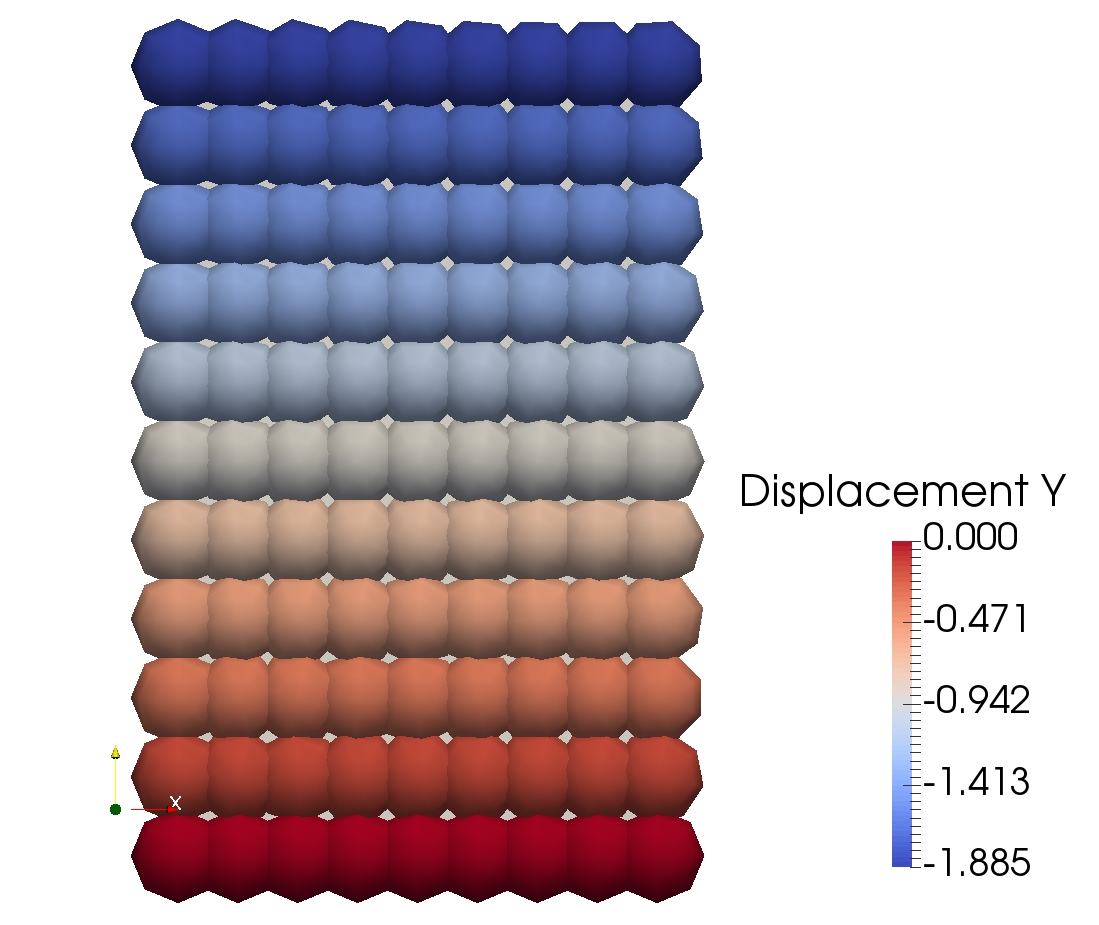}}
		\qquad
		\subfigure[With ML based model]{\label{test1dplas}\includegraphics[width=60mm]{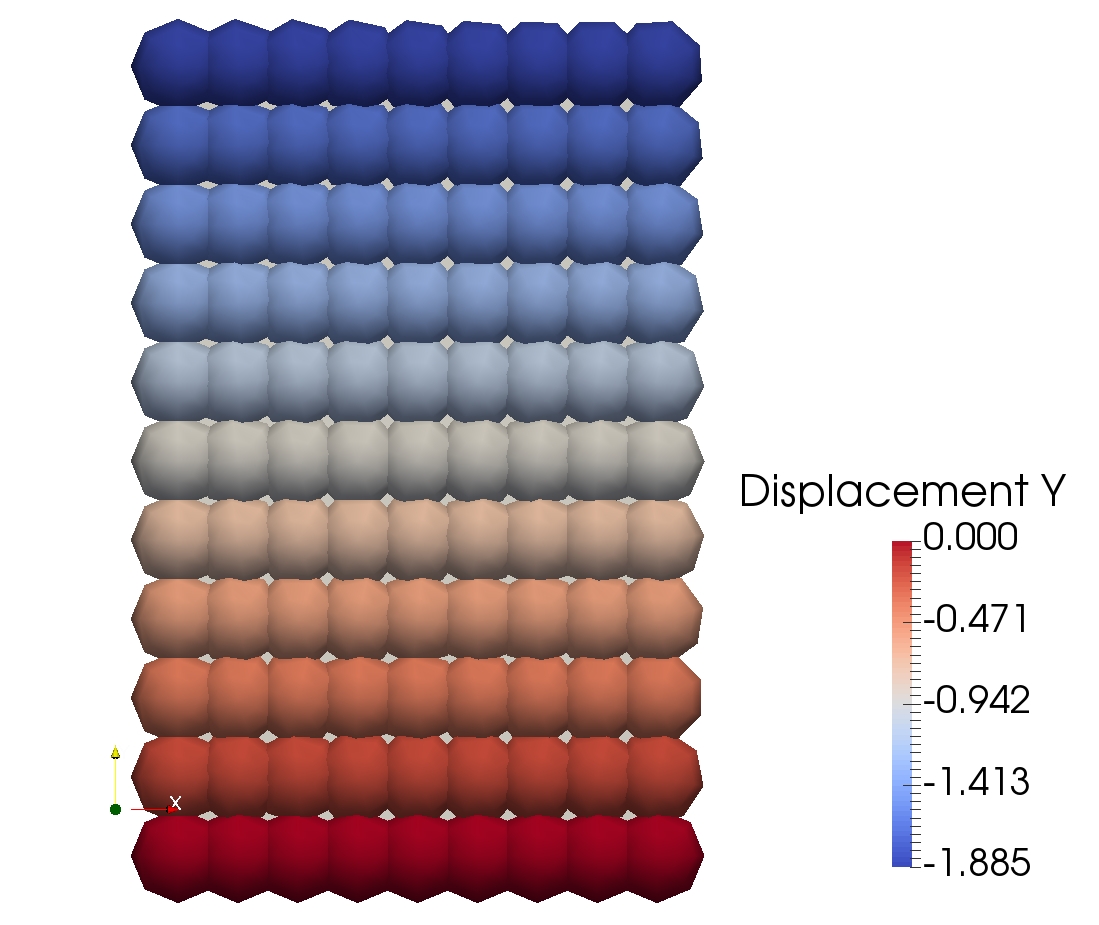}}
		\caption{Deformed state of the plate.}
		\label{barneofnn}
	\end{figure}
	
	In order to further validate the generalization, a second test case, the Cook's membrane problem, is conducted. The tapered beam is clamped at the left end and loaded at the right end by a constant distributed vertical load $q_0=5Mpa$, as depicted in Fig. \ref{cookgeo}. The geometric domain of the structure is discretized by 40 quadratic 9-node quadrilateral elements leading to 189 nodes.\\
	
	\begin{figure}[!htb]
		\centering
		\def\svgwidth{0.4\columnwidth}
		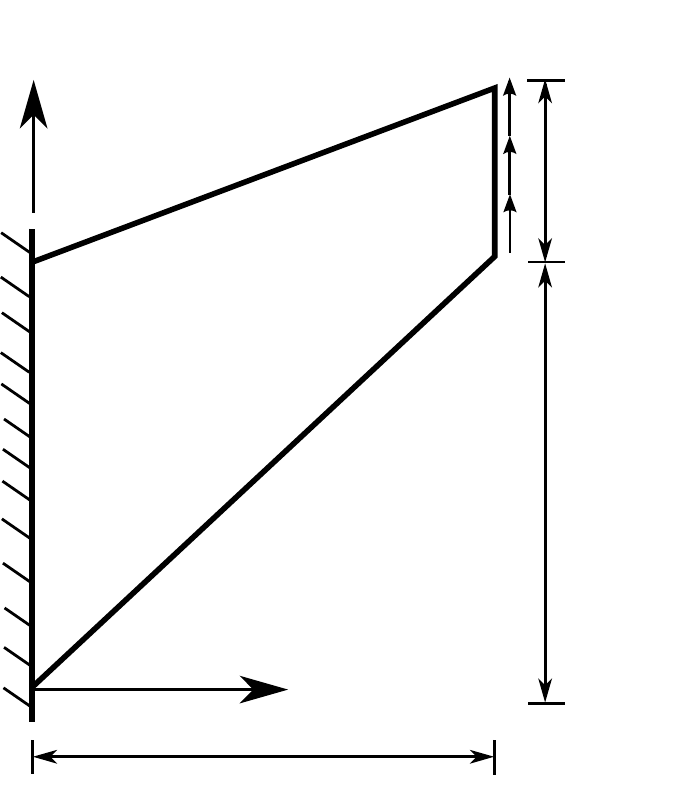
		\caption{Cook's membrane problem.}
		\label{cookgeo}
	\end{figure}
	
	With the same model as trained in the first test, the final deformation of the membrane is computed and compared. As shown in Fig. \ref{cookneofnn}, the vertical displacement in both cases are very close to each other, which highlights the proficient generalization capabilities of the ML based elasticity model. The computation time with analytical hyperelastic model is $15.72s$, whereas the computation time with the ML based model is $19.88s$ on the same computer.
	
	\begin{figure}[!htb]
		\centering
		\subfigure[Neo-Hookean model]{\label{test1dplas}\includegraphics[width=70mm]{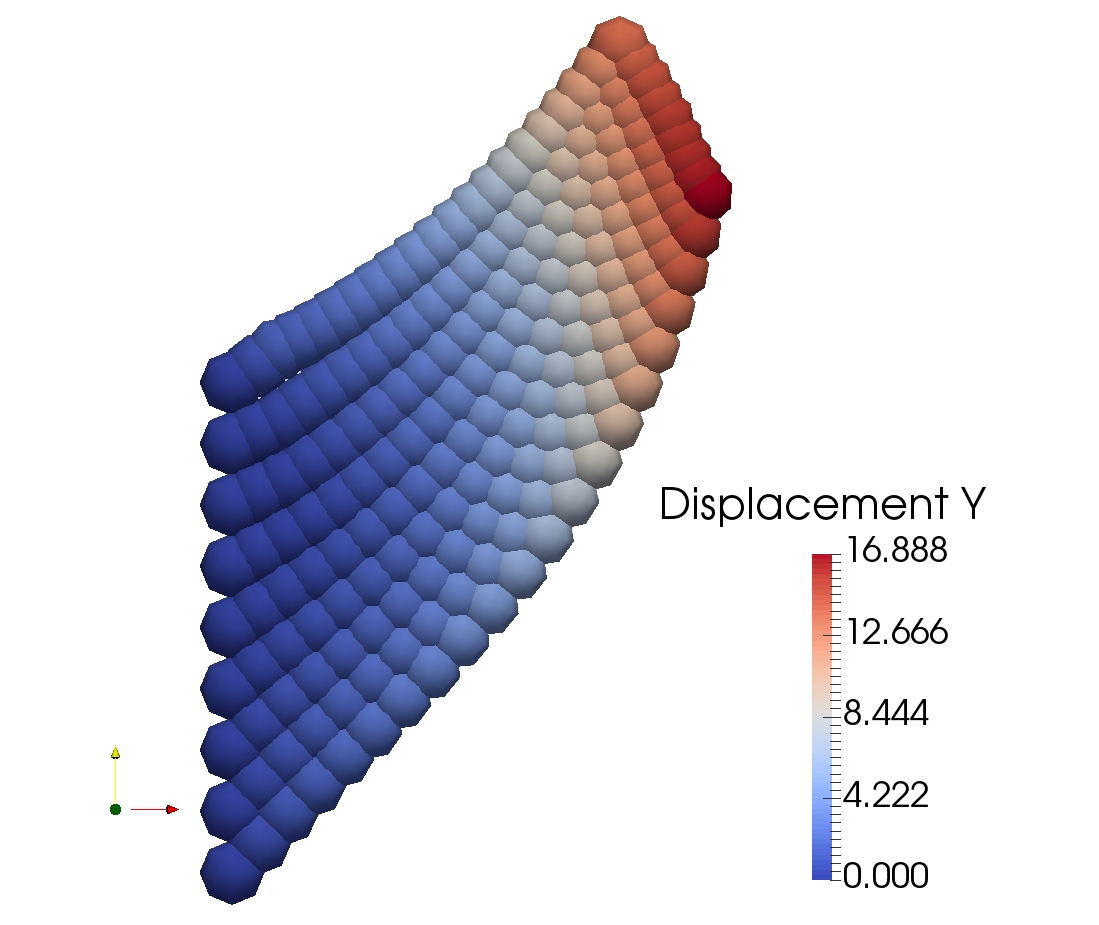}}
		\qquad
		\subfigure[With ML based model]{\label{test1dplas}\includegraphics[width=70mm]{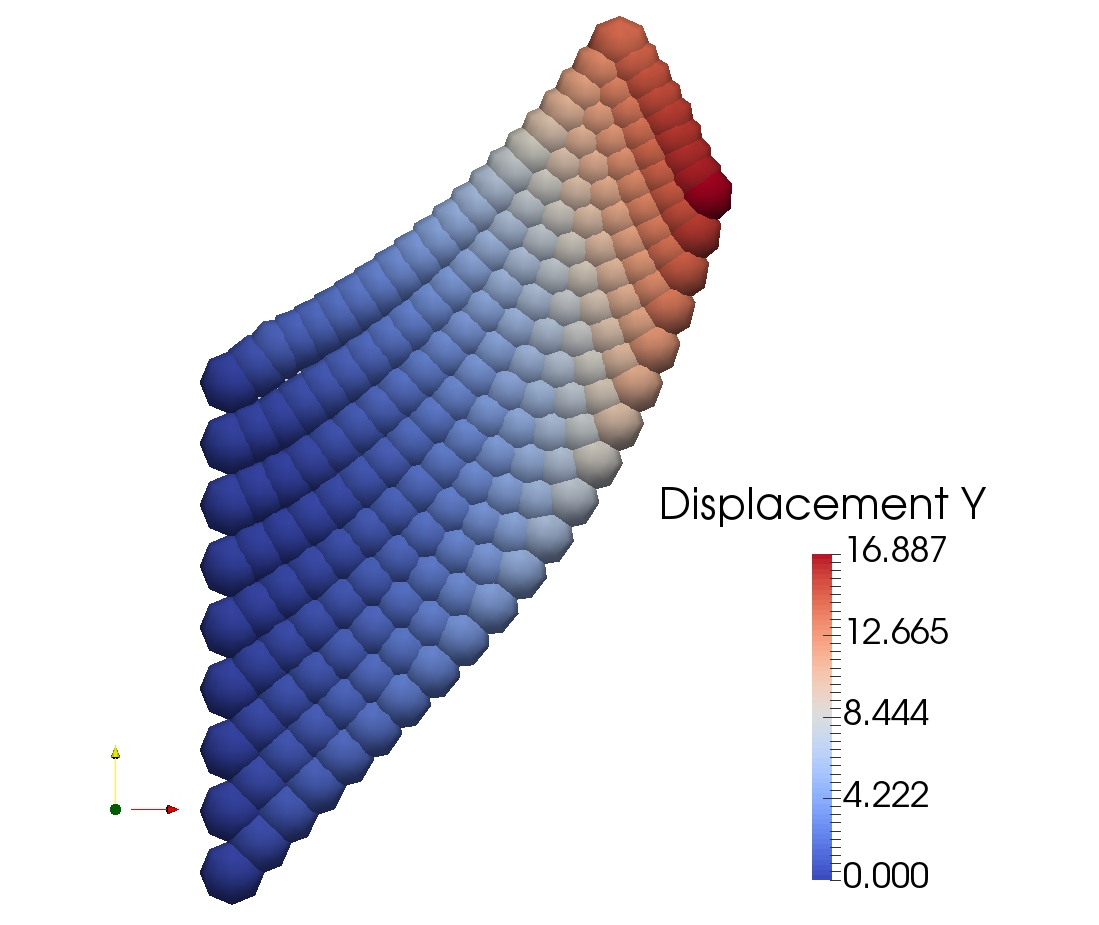}}
		\caption{Deformed state of the Cook's membrane.}
		\label{cookneofnn}
	\end{figure}
	
	To this end, it proves that the FNN works well for approximating the nonlinear elastic behaviour as shown in the above results. Elastic deformation is a history independent process and the stress depends only on current kinematic variables. However, for plastic material behaviour, the response of the deformed material depends not only on the current deformation but also on its loading history. Since the FNN does not have any inherent ability to record loading history, the current approach needs to be improved. Furthermore, the collection process of the training data for plasticity needs to be different from elasticity.
	
	\section{Data collection strategy for plasticity}
	The aim of this part is to develop a data-driven material model which can be used to computationally reproduce the plastic material behavior by means of machine learning tools. Collecting data from experiments is a key part for data-driven material model. In experiments, only the total strain and stress data of a specimen can be collected, which means the classical concept of elastic-plastic splitting to total strain can not be applied in the data-driven model. This leads to the questions: how to build the data-driven model using the total strain and stress data available from experiment? and what kind of experiments have to be conducted to collect data?
	
	\subsection{Concept of sequence for plasticity}
	Since the plastic flow depends not only on the current stress state but also on the loading history, the plastic deformation is a history dependent process. For 1D plastic deformations, the loading and unloading curves do not coincide as shown in Fig. \ref{series1dplas}, where the strain and stress data are time series of data sets and can be seen as sequences.\\
	
	\begin{figure}[!htb]
		\centering
		\def\svgwidth{0.5\columnwidth}
		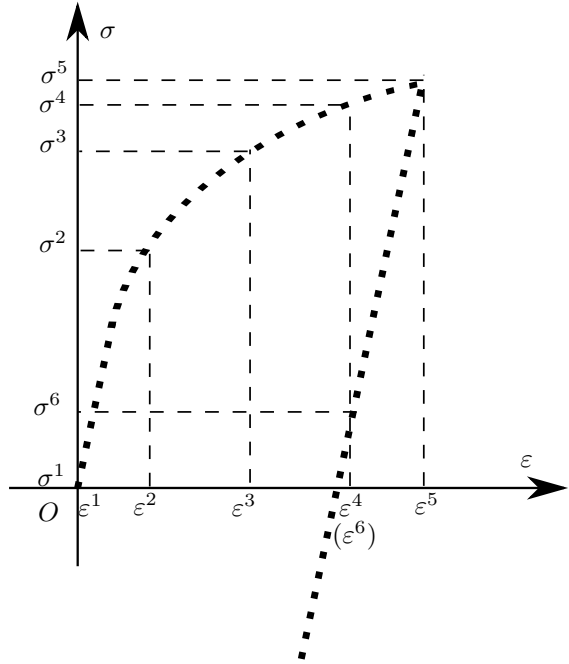
		\caption{The loading and unloading curve for plasticity.}
		\label{series1dplas}
	\end{figure}
	
	Along a 1D loading-unloading path, the strain and stress data sets of one material point have a strict sequential order and can be written as two sets of corresponding sequences
	\begin{align}
	\lbrace \varepsilon^1, \varepsilon^2, \varepsilon^3, ..., \varepsilon^t, ... \rbrace \Leftrightarrow \lbrace \sigma^1, \sigma^2, \sigma^3, ..., \sigma^t, ... \rbrace,
	\end{align}
	where $\varepsilon^t$ and $\sigma^t$ are the total strain and stress at time $t$ collected from the experiment. Thus, the basic data unit for a plasticity model is not a strain-stress pair but a strain-stress sequence pair. Each strain-stress sequence pair refers to one loading-unloading path and thus sequence data collected from different loading-unloading paths are required to train a data-driven plasticity model.
	
	
	In machine learning, the classical equation of plasticity will be replaced with a ML based plasticity model driven by experimental data as
	\begin{align}
		\boldsymbol\sigma^t=f^{ML}(\boldsymbol\varepsilon^t,\boldsymbol h^t),
	\end{align}
	where $\boldsymbol h^t$ is a history variable for distinguishing loading history and $\boldsymbol\varepsilon^t$ is the total strain. In this data-driven model, both the input and the output are sequence data. To build a ML based plasticity model, the history data as well as the strain-stress data have to be obtained from experiments.
	
	The choice of history variable is crucial for a successful prediction of sequences. In the literature, the stress and strain in the last step are applied as the history information together with the current strain in the input, see \cite{hashash2004numerical}. However, the previous strain and previous stress are not enough to distinguish the loading history in real applications, since any error in the predicted previous stress by the ML tool will introduce an extra error into the system in an accumulate way. In this work, the accumulated absolute strain is applied in the input as history variable. The accumulated absolute strain component $\varepsilon_{acc,j}^t$ at time step $t$ can be computed for the $j$-th strain component as
	\begin{align}
	h_j^t:=\varepsilon_{acc,j}^t=
	\begin{cases}
	\sum_{i=3}^{t} \lvert \varepsilon^{i-1}_j -  \varepsilon^{i-2}_j \rvert, & \text{ $t\geqslant3$},\\
	0, & \text{$t=1,2$},
	\end{cases}
	\label{acce}
	\end{align}
	where $\varepsilon^{i-1}_j$ and $\varepsilon^{i-2}_j$ are total strain components at time step $i-1$ and $i-2$ respectively. This history variable has to be computed for each total strain component. $\lvert  \varepsilon^{i-1}_j -  \varepsilon^{i-2}_j \rvert$ is the absolute increment of strain component $j$ from time step $i-2$ to $i-1$, which is necessary to distinguish the loading history when tension and compression loadings are mixed, such as in loading-unloading path.

	For the 1D case, depicted in Fig. \ref{series1dplas}, a monotonic loading (e.g. from $\sigma^1$ to $\sigma^4$) and a mixed loading-unloading (e.g. from $\sigma^1$ to $\sigma^6$) may lead to the same total strain (e.g. $\varepsilon^4=\varepsilon^6$). To distinguish monotonic loading from mixed loading-unloading, the absolute value of strain increment $\lvert \varepsilon^{i-1} -  \varepsilon^{i-2} \rvert$ is applied in equation (\ref{acce}), which leads to different accumulated absolute strain for different paths, e.g.
	\begin{align}
	\varepsilon_{acc}^6&=\sum_{i=3}^{6} \lvert \varepsilon^{i-1} -  \varepsilon^{i-2} \rvert\\
	&=\sum_{i=3}^{4} \lvert \varepsilon^{i-1} -  \varepsilon^{i-2} \rvert+\sum_{i=5}^{6}\lvert \varepsilon^{i-1} -  \varepsilon^{i-2} \rvert\\
	&= \varepsilon_{acc}^4 +\sum_{i=5}^{6}\lvert \varepsilon^{i-1} -  \varepsilon^{i-2} \rvert>\varepsilon_{acc}^4.
	\end{align}
	Note that $ \lvert \varepsilon^{i-1} -  \varepsilon^{i-2} \rvert $ is applied instead of $ \lvert \varepsilon^{i} -  \varepsilon^{i-1} \rvert $ in equation (\ref{acce}), since $\sum_{i=3}^{t} \lvert \varepsilon^{i} -  \varepsilon^{i-1} \rvert$ is equal to $\varepsilon^{t}$ for monotonic loading, and it is not a history variable but the current input. The advantage of applying the accumulated absolute strain as the history variable is that it can be obtained from the existing experimental data without the effort to collect them additionally.
	
	\subsection{Loading paths to collect data from experiments}
	To collect the strain-stress sequence data from experiments, the loading paths required in experiments have to be investigated. Since isotropic plasticity can be formulated in the principle strain-stress space, the ML based plasticity model can be formulated in the principle space as well, where the input and output of the model are principle stain and principle stress respectively. Therefore only the principle strain and principle stress are required to be collected from the experiments.

	To collect the principle strain and principle stresses, the multi-axial loading tests can be conducted on special designed specimens, such as the biaxial experiments conducted by \cite{mohr2010evaluation} and the loading paths suggested by \cite{goel2011finite}. The von Mises yield surface covering different stress states is shown in Fig. \ref{fig:yieldsurf} for the 2D case. In order to learn the plasticity behavior by e.g. artificial neural networks, yielding as well as hardening effects have to be captured implicitly. To fully describe the stress states existed in the deformed structures, the data have to be collected from several tests under different loading-unloading paths. However, only biaxial tests for 2D and triaxial tests for 3D are required to collect data in principle space.\\
	
	\begin{figure}[!htb]
		\centering
		\def\svgwidth{0.6\columnwidth}
		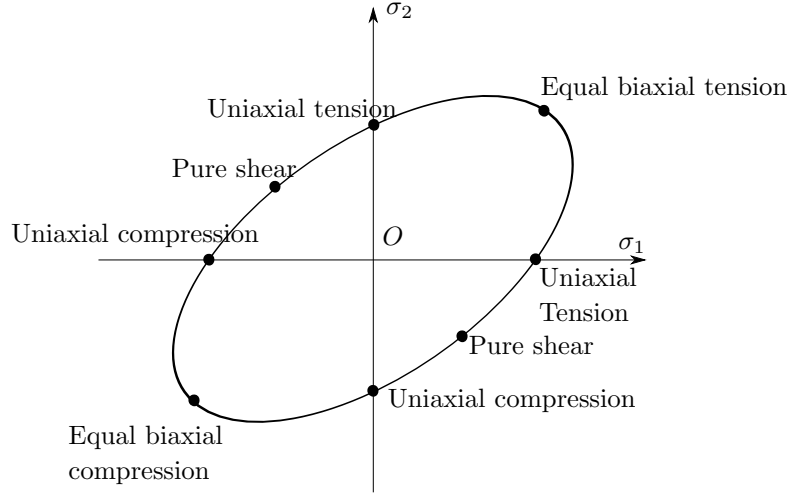
		\caption{Stress states in 2D.}
		\label{fig:yieldsurf}
	\end{figure}

In experiments, the principle strain and stress data can be collected within a homogeneous region at one point within a specimen, which can be descried by a quadrilateral region for 2D case depicted in Fig. \ref{ele4}. The biaxial loading-unloading paths at this region can be characterized by the displacements of the edges connected to point $A$.\\

	\begin{figure}[!htb]
		\centering
		\def\svgwidth{0.4\columnwidth}
		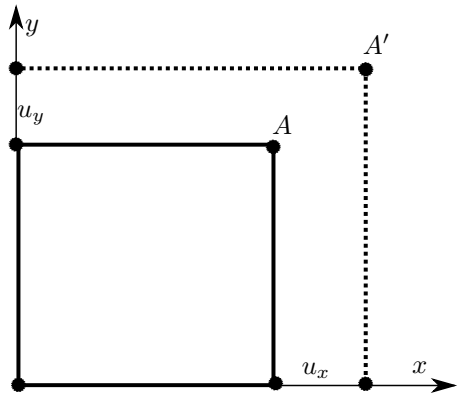
		\caption{Quadrilateral region within a specimen to collect data with biaxial loading.}
		\label{ele4}
	\end{figure}
	
For each biaxial loading-unloading path, the point $A$ moves from its original position to $A'$ for loading and then goes back to original position for unloading, during which the displacements $(u_x, u_y)$ will increase linearly from zero to a specific value obeying $\sqrt{u_x^2+u_y^2}=r_i, (i=1,2,3...)$ and then decrease to zero. $max(r_i)$ has to be large enough to capture as much loading range of plastic deformation as possible. As shown in Fig. \ref{path-circle}, the loading-unloading paths are just determined by setting a radius $r_i$ and different values of the angles $\phi$. Since the unloading can start from different positions, multiple circles with radius $r_i$ have to be applied in data collection.\\
	
	\begin{figure}[!htb]
		\centering
		\def\svgwidth{0.56\columnwidth}
		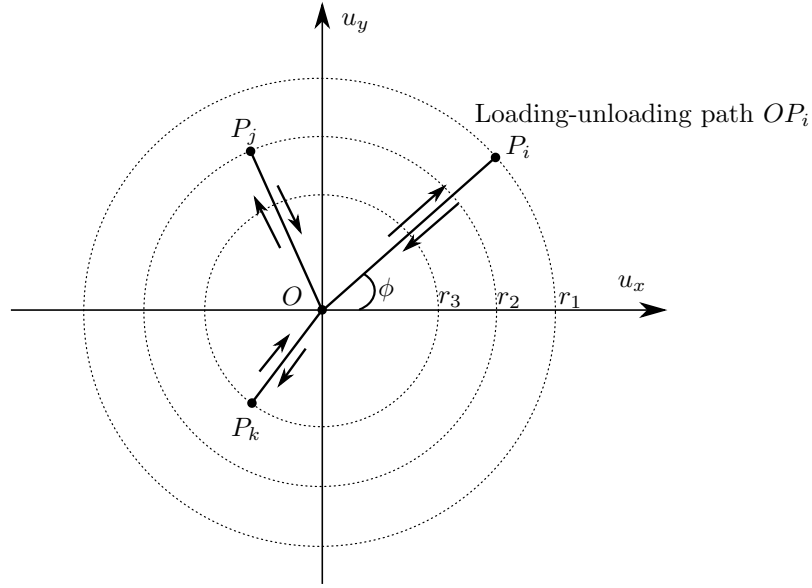
		\caption{Loading-unloading paths for data collection in 2D.}
		\label{path-circle}
	\end{figure}
		
For the 3D case, data can be collected from a cubic region, as shown in Fig. \ref{ele8}, where the triaxial loading-unloading paths at this point can be characterized by the displacement $(u_x, u_y, u_z)$ at the point $A$. \\
	
	\begin{figure}[!htb]
		\centering
		\def\svgwidth{0.4\columnwidth}
		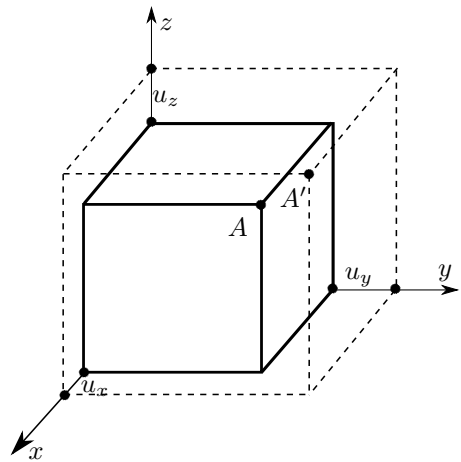
		\caption{Cubic region within a specimen to collect data with triaxial loading.}
		\label{ele8}
	\end{figure}
The loading-unloading path for the 3D case can be generated in the spherical coordinate system as shown in Fig. \ref{path-sphere}, where the displacement of point $A$ obeys $\sqrt{u_x^2+u_y^2+u_z^2}=r_i, (i=1,2,3,...)$. The loading path $OP_i$ is distinguished by the angles $\phi$ and $\theta$ with radius $r_i$. By looping the loading path around the sphere, the whole range of stress states can be included.\\
	
	\begin{figure}[!htb]
		\centering
		\def\svgwidth{0.5\columnwidth}
		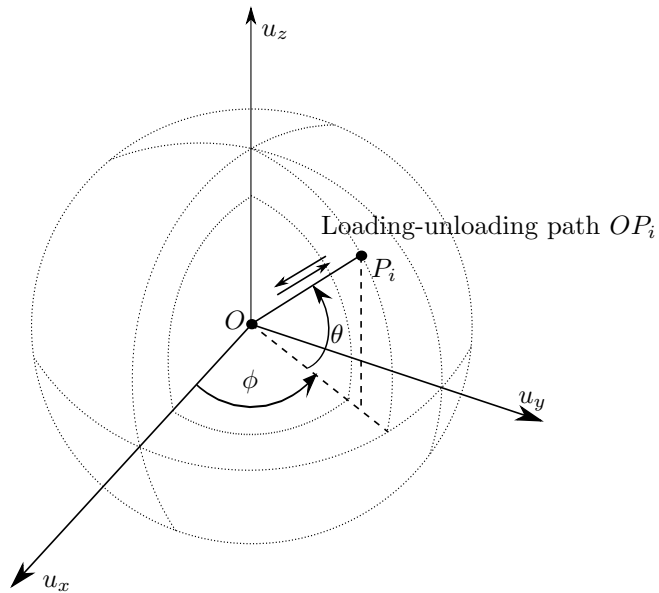
		\caption{Loading-unloading paths for data collection in 3D.}
		\label{path-sphere}
	\end{figure}

For example, along the loading path $OP_i$ in Fig. \ref{path-sphere}, the strain-stress sequence data will be collected firstly as
\begin{align}
\hat{\boldsymbol\varepsilon} =	\begin{bmatrix}
\varepsilon^1_1 & \varepsilon^2_1 & ... & \varepsilon^t_1 & ...\\
\varepsilon^1_2 & \varepsilon^2_2 & ... & \varepsilon^t_2 & ...\\
\varepsilon^1_3 & \varepsilon^2_3 & ... & \varepsilon^t_3 & ...\\
\end{bmatrix}_{3 \times np}, \,\,\,\,\, 
\boldsymbol\sigma_{opi} =	\begin{bmatrix}
\sigma^1_1 & \sigma^2_1 & ... & \sigma^t_1 & ...\\
\sigma^1_2 & \sigma^2_2 & ... & \sigma^t_2 & ...\\
\sigma^1_3 & \sigma^2_3 & ... & \sigma^t_3 & ...\\
\end{bmatrix}_{3 \times np},
\end{align}
where $np$ is the number of data point on the loading path $OP_i$, $ \varepsilon^t_i (i=1,2,3)$ are the principle strains at time $t$ measured in the cubic region within the specimen, $ \sigma^t_i (i=1,2,3)$ are the principle stresses at time $t$ in that region, $\hat{\boldsymbol\varepsilon}$ is the strain sequence and $\boldsymbol\sigma_{opi}$ is the stress sequence of path $OP_i$. Then the history data, accumulated absolute strain $\varepsilon_{acc}^t$, are computed from the strain sequence $\hat{\boldsymbol\varepsilon}$ using equation (\ref{acce}). The final strain sequence data of path $OP_i$ are obtained by combining the total strain sequence and the history variable sequence as
\begin{align}
\boldsymbol\varepsilon_{opi}=	\begin{bmatrix}
\varepsilon^1_1 & \varepsilon^2_1 & ... & \varepsilon^t_1 & ...\\
\varepsilon^1_2 & \varepsilon^2_2 & ... & \varepsilon^t_2 & ...\\
\varepsilon^1_3 & \varepsilon^2_3 & ... & \varepsilon^t_3 & ...\\
\varepsilon_{acc,1}^1 & \varepsilon_{acc,1}^2 & ... & \varepsilon_{acc,1}^t & ...\\
\varepsilon_{acc,2}^1 & \varepsilon_{acc,2}^2 & ... & \varepsilon_{acc,2}^t & ...\\
\varepsilon_{acc,3}^1 & \varepsilon_{acc,3}^2 & ... & \varepsilon_{acc,3}^t & ...\\
\end{bmatrix}_{6 \times np},
\end{align}
where $\varepsilon_{acc,1}^t$, $\varepsilon_{acc,2}^t$ and $\varepsilon_{acc,3}^t$ are computed from the strain components $\varepsilon^t_1$, $\varepsilon^t_2$ and $\varepsilon^t_3$ respectively according to equation (\ref{acce}).
Finally, the input and output data are obtained by combining the sequences from all of loading paths $OP_i, (i=1,2,...,nl)$ as
\begin{align}
\boldsymbol I_{\varepsilon} =	\begin{bmatrix}
\boldsymbol\varepsilon_{op1} & \boldsymbol\varepsilon_{op2} & ... & \boldsymbol\varepsilon_{opi} & ...
\end{bmatrix}_{6 \times M}, \,\,\,\,\, 
\boldsymbol O_{\sigma} =	\begin{bmatrix}
\boldsymbol\sigma_{op1} & \boldsymbol\sigma_{op2} & ... & \boldsymbol\sigma_{opi} & ...
\end{bmatrix}_{3 \times M},
\label{IO}
\end{align}
where $nl$ is the number of loading path and $M=np \times nl$.

\section{Machine learning based plasticity}
After data collection, the feed forward neural network is used to learn the relationship within the data. The neural network will approximate a mapping between inputs and outputs. However, the accuracy of the approximation depends on the complexity of the relationship. As a widely used technique in model order reduction, the Proper Orthogonal Decomposition (POD) provides an approach to decouple the training data, which simplifies the training and increases accuracy.

\subsection{Decouple the stress data by POD}
The Proper Orthogonal Decomposition (POD) in combination with machine-learning tools, such as Gaussian Processes (\cite{xiao2010model}) and Long-Short-Term memory network (\cite{mohan2018deep}), as a reduced order model has been used to surrogate model generation of fluid dynamic systems. Here we introduce a novel combination framework, where POD and FNNs are combined for preprocessing and prediction of sequence data. We call this approach Proper Orthogonal Decomposition Feed forward Neural Network (PODFNN). Using POD, the time series vector variables can be represented with a reduced number of modes neglecting higher order modes if the error is acceptable. Thus, by use of the POD, the problem will be decoupled into a combination of several different modes.

As time series data, the stress sequence in training data can be rewritten as a snapshot matrix
	\begin{align}
	\boldsymbol O_{\sigma}=
	\begin{bmatrix}
	\boldsymbol\sigma_{op1} & \boldsymbol\sigma_{op2} & ... & \boldsymbol\sigma_{opi} & ...
	\end{bmatrix}
	=
	\begin{bmatrix}
	\sigma^1_1 & \sigma^2_1 & ... & \sigma^k_1 & ...\\
	\sigma^1_2 & \sigma^2_2 & ... & \sigma^k_2 & ...\\
	\sigma^1_3 & \sigma^2_3 & ... & \sigma^k_3 & ...
	\end{bmatrix}_{3 \times M},
	\end{align}
	 where each column of the matrix is a snapshot and can be written as a vector $\boldsymbol o_{\sigma}^k =\begin{bmatrix} \sigma^k_1 & \sigma^k_2 & \sigma^k_3 \end{bmatrix}^T$. Using POD, any snapshot $\boldsymbol o_{\sigma}^k$ can be represented by a linear combination of the basis
	\begin{align}
	\boldsymbol o_{\sigma}^k = \bar{\boldsymbol \sigma} + \sum_{i=1}^{m} \alpha_i^k \boldsymbol{\varphi}_i,
	\end{align}
	where $\boldsymbol{\varphi}_i=\begin{bmatrix} \phi^k_1 & \phi^k_2 & \phi^k_3 \end{bmatrix}^T$ is the $i$-th basis vector, $\alpha^k_i$ is the coefficient, $m$ is the number of POD mode and $\bar{\boldsymbol \sigma}=\begin{bmatrix}\bar{\sigma}_1 & \bar{\sigma}_2 & \bar{\sigma}_3\end{bmatrix}^T$ is the mean value vector of the snapshot matrix. Since $\bar{\boldsymbol \sigma}$, $\boldsymbol{\varphi}_i$ are constants, and the bases $\boldsymbol{\varphi}_i$ are independent with each other, the stress sequence can thus be decoupled into independent coefficient sequences.
	
	The components of mean value vector $\bar{\boldsymbol \sigma}$ of the snapshot matrix are computed as
	\begin{align}
	\bar{\sigma}_1=\frac{1}{M}\sum_{i=1}^{M} \sigma^i_1, \,\,\,\,\,\, \bar{\sigma}_2=\frac{1}{M}\sum_{i=1}^{M} \sigma^i_2, \,\,\,\,\,\, \bar{\sigma}_3=\frac{1}{M}\sum_{i=1}^{M} \sigma^i_3.
	\end{align}
	To find the basis vectors and its coefficients, the deviation matrix is firstly computed as
	\begin{align}
	\boldsymbol O_{\sigma}^{dev}=\begin{bmatrix}
	\sigma^1_1-\bar{\sigma}_1 & \sigma^2_1-\bar{\sigma}_1 & ... & \sigma^k_1-\bar{\sigma}_1 & ...\\
	\sigma^1_2-\bar{\sigma}_2 & \sigma^2_2-\bar{\sigma}_2 & ... & \sigma^k_2-\bar{\sigma}_2 & ...\\
	\sigma^1_3-\bar{\sigma}_3 & \sigma^2_3-\bar{\sigma}_3 & ... & \sigma^k_3-\bar{\sigma}_3 & ...
	\end{bmatrix}_{3 \times M}.
	\end{align}
	Then, by applying Singular Value Decomposition (SVD) to the deviation matrix
	\begin{align}
	\boldsymbol O_{\sigma}^{dev}=\boldsymbol U \boldsymbol S \boldsymbol V^T,
	\end{align}
	where $\boldsymbol U$ and $\boldsymbol V$ are the unitary matrices, $\boldsymbol S$ is the diagonal matrix with non-negative real numbers on the diagonal, the basis vectors $\boldsymbol{\varphi}_i$ can be determined from the non-zero columns of matrix $\boldsymbol U$
	\begin{align}
	\boldsymbol\Phi=\begin{bmatrix}\boldsymbol{\varphi}_1 & \boldsymbol{\varphi}_2 & ... &\boldsymbol{\varphi}_m \end{bmatrix}=\begin{bmatrix} \boldsymbol u_1 & \boldsymbol u_2 & ... & \boldsymbol u_m \end{bmatrix}_{3 \times m},
	\end{align}
	where $\boldsymbol u_m$ is the $m$-th non-zero column of matrix $\boldsymbol U$ and $m$ is equal to the rank of $\boldsymbol O_{\sigma}^{dev}$.
	
	The coefficients $\boldsymbol \alpha^k=\begin{bmatrix} \alpha^k_1 & \alpha^k_2& ...& \alpha^k_m \end{bmatrix}^T$ can be obtained by projecting the snapshot $\boldsymbol o^k_{\sigma}$ on the basis matrix $\boldsymbol \Phi$
	\begin{align}
	\boldsymbol \alpha^k=\boldsymbol \Phi^T\boldsymbol o^k_{\sigma}.
	\end{align}
	
	Since the stress components are independent, the rank of matrix $\boldsymbol O_{\sigma}^{dev}$ is $3$ in this work ($m=3$). The stress sequence in equation (\ref{IO}) will be represented by $3$ independent coefficient sequences
	\begin{align}
	\boldsymbol O_{\sigma}=
	\begin{bmatrix}
	\sigma^1_1 & \sigma^2_1 & ... & \sigma^k_1 & ...\\
	\sigma^1_2 & \sigma^2_2 & ... & \sigma^k_2 & ...\\
	\sigma^1_3 & \sigma^2_3 & ... & \sigma^k_3 & ...
	\end{bmatrix}_{3 \times M}
	\frac{POD}{\rightarrow}
	\begin{array}{lcr}
	\begin{bmatrix} \alpha^1_1 & \alpha^2_1 & ... & \alpha^k_1& ... \end{bmatrix}_{1 \times M}\\
	\begin{bmatrix} \alpha^1_2 & \alpha^2_2 & ... & \alpha^k_2& ... \end{bmatrix}_{1 \times M}\\
	\begin{bmatrix} \alpha^1_3 & \alpha^2_3 & ... & \alpha^k_3& ... \end{bmatrix}_{1 \times M},\end{array}.
	\end{align}
	Therefore, the training data composed by the strain-stress sequences in equation (\ref{IO}) for plasticity model is transformed into training data composed by strain-coefficient sequences and can be written as
	\begin{align}
	\begin{bmatrix}
	\boldsymbol\varepsilon_{op1} & \boldsymbol\varepsilon_{op2} & ... & \boldsymbol\varepsilon_{opi} & ...
	\end{bmatrix}_{6 \times M} \Leftrightarrow 
	\begin{array}{lcr}
	\begin{bmatrix} \alpha^1_1 & \alpha^2_1 & ... & \alpha^k_1& ... \end{bmatrix}_{1 \times M}\\
	\begin{bmatrix} \alpha^1_2 & \alpha^2_2 & ... & \alpha^k_2& ... \end{bmatrix}_{1 \times M}\\
	\begin{bmatrix} \alpha^1_3 & \alpha^2_3 & ... & \alpha^k_3& ... \end{bmatrix}_{1 \times M},\end{array}
	\label{mapping}
	\end{align}
	where the three coefficient sequences are independent from each other.
				
	\subsection{Prediction of coefficients using FNN}
	Once the training data is prepared, FNNs are applied to learn the mapping between the strain sequence and the coefficient sequences in equation (\ref{mapping}). Since the coefficients in the POD representation are independent, each coefficient can be predicted by one FNN, as shown in Fig. \ref{PODFNN}, where the original strain-stress mapping approximated by one complex neural network is decoupled into three independent stain-coefficient mappings approximated by simpler neural networks.\\
	
	\begin{figure}[!htb]
		\centering
		\def\svgwidth{0.95\columnwidth}
		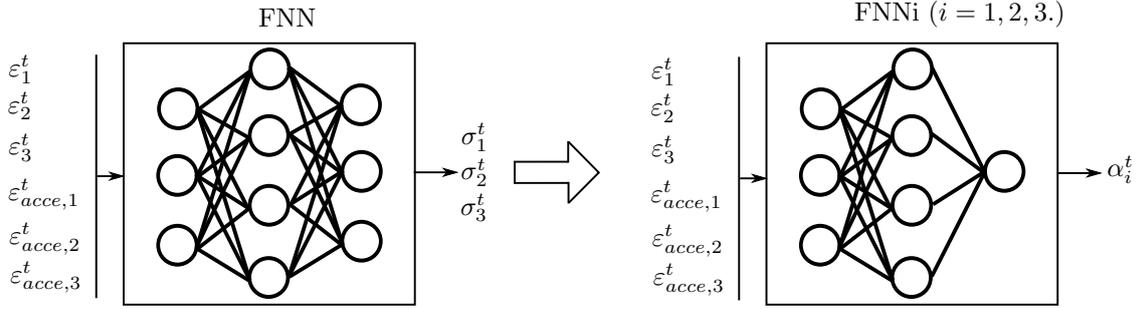
		\caption{Decoupling the strain-stress mapping (left) into independent strain-coefficient mappings (right) by POD.}
		\label{PODFNN}
	\end{figure}
	
	After training, the coefficient $\alpha^t_{NN,i}$ at time $t$ will be predicted by the feed forward neural network $FNN_i$ as
	\begin{align}
	\alpha^t_{NN,i} = FNN_i (\boldsymbol \varepsilon^t,\boldsymbol \varepsilon_{acc}^t, \boldsymbol w, \boldsymbol b_s), \,\,\, (i=1,2,3),
	\label{aFNN}
	\end{align}
	where $i$ indicates the number of the coefficient, $\boldsymbol \varepsilon^t=\begin{bmatrix} \varepsilon_1^t &\varepsilon_2^t & \varepsilon_3^t
	\end{bmatrix}^T$ is the current strain, $\boldsymbol \varepsilon_{acc}^t=\begin{bmatrix} \varepsilon_{acc,1}^t&\varepsilon_{acc,2}^t&\varepsilon_{acc,3}^t\end{bmatrix}^T$ is the accumulated absolute strain, $\boldsymbol w$ is the weight matrix and $\boldsymbol b_s$ is the bias of neural network.
		
	\subsection{PODFNN based plasticity model}
	Once the coefficient $\boldsymbol\alpha^t_{NN}$ is predicted by FNNs as described in equation (\ref{aFNN}), the principle stress can be recovered from the POD representation as
	\begin{align}
	\tilde{\boldsymbol \sigma}^t_{PODFNN} = \bar{\boldsymbol \sigma} + \sum_{i=1}^{3} \alpha_{NN,i}^t \boldsymbol{\varphi}_i.
	\end{align}
	Finally, the Cauchy stress is obtained by transforming the principle stress into the general space
	\begin{align}
	\boldsymbol\sigma^t_{PODFNN} =\boldsymbol{Q}\tilde{\boldsymbol \sigma}^t_{PODFNN} \boldsymbol{Q}^T.
	\end{align}
	The formulation of the ML based plasticity model defines a constitutive function in the following format
	\begin{align}
	\boldsymbol \sigma_{PODFNN}^t=PODFNN (\boldsymbol \varepsilon_t, \boldsymbol{\varepsilon}_{acc}^t, \boldsymbol{w}, \boldsymbol{b}_s),
	\end{align}
	where $\boldsymbol \varepsilon_t$ is the current total strain, $\boldsymbol{\varepsilon}_{acc}^t$ is the history variable.
	To apply the ML based plasticity model in the finite element analysis, the residual vector and tangent matrix have to be derived. The automatic symbolic differentiation tool AceGen is applied to derive the tangent matrix again.
	
	\subsection{Testing the ML based plasticity model in FEM}
	In the following, the performance of the developed ML based plasticity model is evaluated using finite element applications.
	\subsubsection{Data collection from analytical model}
	Apart from collecting the training data from experiments, simulation data using the von Mises plasticity model can be collected as well to train the ML tool, in this way the performance of the ML based plasticity model can be verified by comparing to the analytical model. In this work, the strain-stress sequences are collected at a Gauss point of a finite element under specific loading paths described above.
	The strain is computed as the symmetric part of the displacement gradient for small deformations
	\begin{align}
	\boldsymbol\varepsilon=\frac{1}{2}(\boldsymbol{H}+\boldsymbol{H}^T),
	\end{align}
	where $\boldsymbol{H}$ is the displacement gradient with $\boldsymbol{H}={\rm Grad}\boldsymbol{u}$. Using spectral decomposition, the strain can be formulated as
	\begin{align}
	\boldsymbol\varepsilon =\boldsymbol Q \cdot \boldsymbol\Lambda \cdot \boldsymbol Q^T,
	\label{strain}
	\end{align}
	where $\boldsymbol\Lambda$ is the principle strain and $\boldsymbol Q$ is the rotation matrix obtained from the eigendirections. The principle strain $\boldsymbol\Lambda$ along different loading paths will serve as input data to train the ML based plasticity model.
	
	For small strain plasticity, the additively decomposition of strain into elastic part and plastic part is assumed
	\begin{align}
	\boldsymbol\Lambda =\boldsymbol\Lambda^e +\boldsymbol\Lambda^p,
	\end{align}
	where $\boldsymbol\Lambda^e$ and $\boldsymbol\Lambda^p$ are the elastic strain and plastic strain respectively. The plastically admissible stress is given by
	\begin{align}
	\boldsymbol \Sigma=\rho \frac{\partial \psi}{\partial \boldsymbol\Lambda^e},
	\end{align}
	where $\psi$ is the free energy function. The principle stress $\boldsymbol \Sigma$ will serve as the output data, corresponding to the principle strain as input. By assuming the von Mises yield criteria, the yield function is written as
	\begin{align}
	f=\sqrt{\frac{3}{2}} \Vert\boldsymbol{\boldsymbol \Sigma}^{dev} \Vert -\sigma_y(\alpha),
	\end{align}
	where $\boldsymbol \Sigma^{dev}$ is the deviatoric stress with $\boldsymbol \Sigma^{dev} = \boldsymbol\Sigma - \frac{1}{3} tr\boldsymbol\Sigma \cdot \boldsymbol{1}  $ and $\alpha$ is the isotropic hardening variable. By use of the associated plastic flow rule, the evolution equations for the principle plastic strain and the hardening variable are formulated as
	\begin{align}
	\dot{\boldsymbol\Lambda}^p=\dot{\gamma}\frac{\partial f}{\partial \boldsymbol \Sigma^{dev}}, \,\,\,\,\,  \dot{\alpha}=\dot{\gamma}\frac{\partial f}{\partial A}
	\end{align}
	where $\gamma$ is the plastic multiplier and $A$ is the thermodynamic force conjugate with $\alpha$. The plastic flow has to full fill the Kuhn-Tucker conditions
	\begin{align}
	f \leqslant 0, \,\,\,\,\, \dot{\gamma} \geqslant0,  \,\,\,\,\, \dot{\gamma} f=0.
	\end{align}
	\subsubsection{Uniaxial tension and compression}
	To test the performance of the ML based plasticity model, the 1D uniaxial tension and compression test is conducted firstly. The von Mises plasticity with the linear isotropic hardening is applied as the target model. The material parameters of the plasticity model are set as: Young's modulus $E=700N/mm^2$, yield stress $\sigma_y=100MPa$ and isotropic hardening parameter $H_{iso}=10$. To prepare the training data, 11 sets of strain-stress sequence data are collected from the target model with strain increments being increased linearly from $0.02$ to $0.03$. The stress sequences are then transformed into the coefficient sequence by the POD. 
	
	The feed forward neural network with the architecture of (2-20-20-1) is applied to predict the coefficient, where the input layer containing 2 neurons is connected with two hidden layers containing 20 neurons each. The output layer contains one neuron. The input of the neural network is the total strain together with the accumulated absolute strain, and the output of the neural network is the coefficient transformed from the stress sequence. The Levenberg-Marquardt algorithm is applied as the optimizer in training. The weights are initialized by the Nguyen-Widrow method. After 4000 epochs, the mean squared error decreased to $0.0393$ which costs the training time of 2m24s.
	
	The testing strain sequence is generated by setting the strain increment as 0.15 so that it is different from the training data. The stress computed from the ML based plasticity model is compared with the the stress from the target plasticity model, as shown in Fig. \ref{POD1D}. It can be seen that the predicted stress follows the exact solution well, which validates the accuracy of the proposed machine learning approach for plasticity. This test shows that the accumulated absolute total strain as a history variable captures the loading history for the cyclic loading condition.\\
	
	\begin{figure}[!htb]
		\centering
		\def\svgwidth{0.6\columnwidth}
		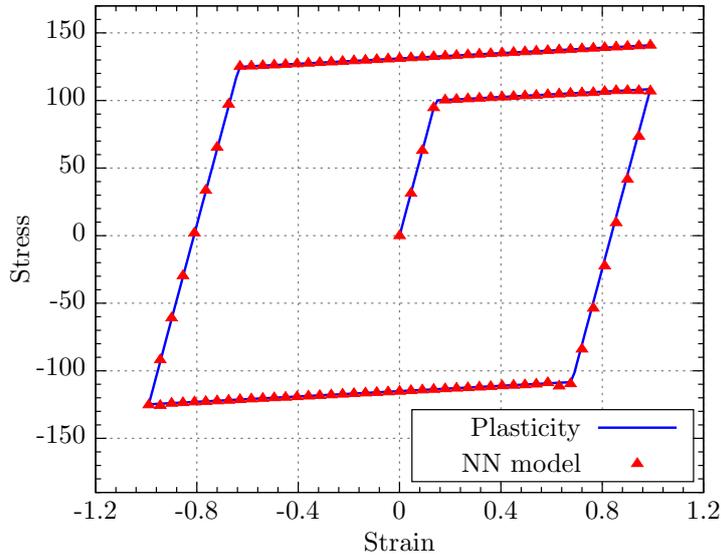
		\caption{Uniaxial tension and compression.}
		\label{POD1D}
	\end{figure}
	
	\subsubsection{Applications in 2D finite element analysis}
	To evaluate the ML based plasticity model, benchmark tests in 2D are presented. The von Mises plasticity with an exponential isotropic hardening law $\sigma_y = y_0+y_0(0.00002+\gamma)^{0.3}$ is set as the target model. The material parameters are set as: Young's modulus $E=1N/mm^2$, Poison's ratio $\nu=0.33$, and the initial yield stress $y_0=0.05MPa$. To collect the training data, $122$ loading-unloading paths evenly distributed within the circles ($r_1=0.1, r_2=0.075$) in Fig. \ref{path-circle} are selected to conduct the biaxial tests, where $61$ values are assigned to the angle $\phi$. Then the collected stress sequence data is transformed into the coefficient sequences using POD.
	
	Since there are two coefficients referring to the two principle stress components in the 2D case, two FNNs will be required to predict the coefficients. In this part, the same network architecture (4-20-20-1) is applied for the two FNNs, where the input layer containing 4 neurons is connected with two hidden layers containing 20 neurons each. The output layer contains always one neuron. The total strain together with the accumulated absolute strain are applied as the input of the neural network. The output of the neural network is the coefficient transformed from the stress sequences. The Levenberg-Marquardt algorithm is applied as the optimizer as well. The training progress is terminated when the gradient of error is less than $10^{-7}$, where the mean squared error is decreased to $7.96 \times 10^{-9}$ for the first FNN and $6.35 \times 10^{-9}$ for the second FNN. 
	
	After the training process, the weighs and biases of the neural network are output as the constant model parameters, by which the Cauchy stress is recovered according to the POD formulation. The tangent matrix and the residual vector are derived using the symbolic differentiation tool AceGen again.
	
	Firstly, the 2D ML based plasticity model is tested by the Cook's membrane problem. The beam is clamped at the left end and loaded at the right end by a constant distributed vertical load $q_0=0.03Mpa$, as depicted in Fig. \ref{cookgeo}. In the unloading process, the direction of vertical load is changed to be negative. The geometric domain of the structure is discretized by 40 quadratic 9-node quadrilateral elements leading to 189 nodes.
	
	Before unloading, the final deformation state of the beam using the proposed ML based plasticity model is compared with the target plasticity model, which is depicted in Fig. \ref{cook2D}. It can be observed that the vertical displacement of the structure is almost identical.\\
	
	\begin{figure}[!htb]
		\centering
		\subfigure[With plasticity model]{\label{cookj2}\includegraphics[width=75mm]{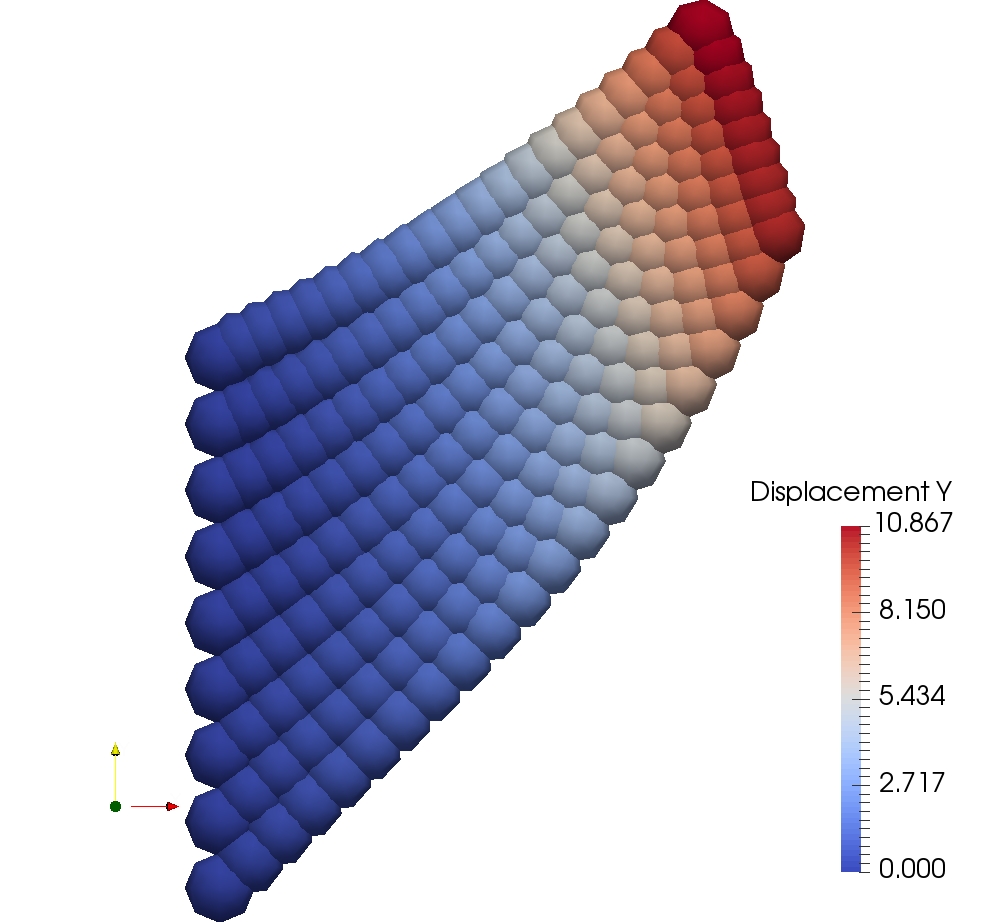}}
		\subfigure[With NN model]{\label{cookrnn}\includegraphics[width=75mm]{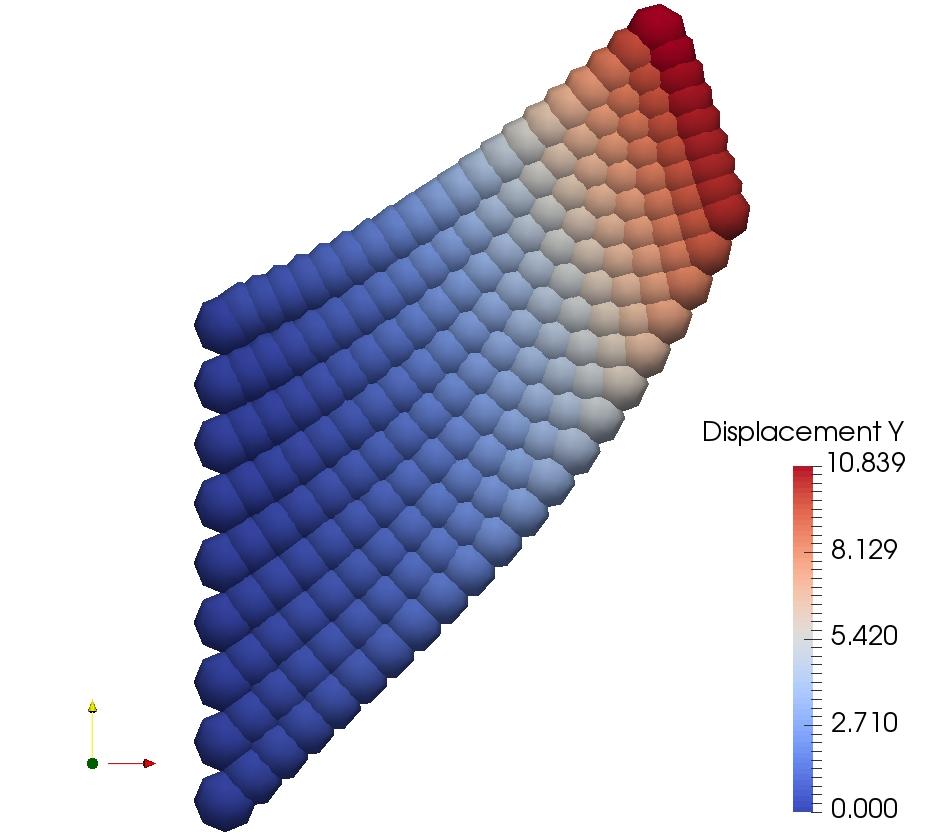}}
		\caption{Final deformation state of the 2D Cook's membrane.}
		\label{cook2D}
	\end{figure}
	The load displacement curve of the upper node $(48, 60)$ at the right end of the cantilever beam is plotted in Fig. \ref{cookUL}. The figure shows that the ML based plasticity model captures the loading and unloading behaviour very well. 
	
	\begin{figure}[!htb]
		\centering
		\def\svgwidth{0.6\columnwidth}
		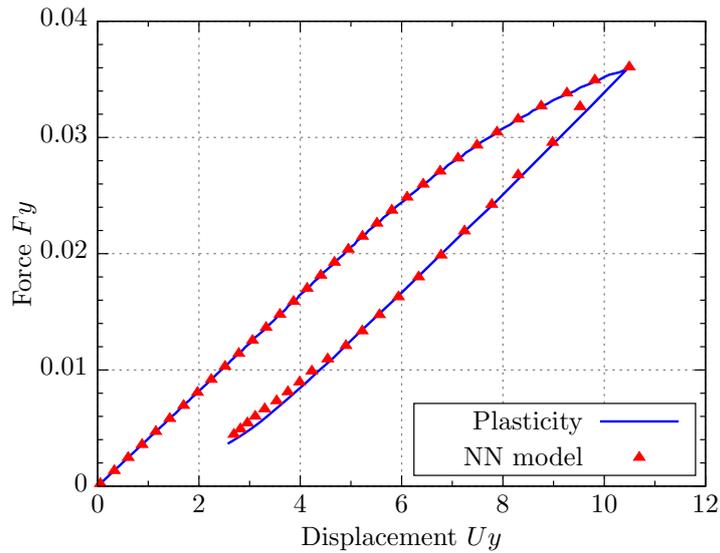
		\caption{Load deflection curve of the 2D Cook's membrane.}
		\label{cookUL}
	\end{figure}

The second example to test the ML based plasticity model is a punch test as shown in Fig. \ref{punch2Dgeo}, where the vertical displacement boundary condition ($u_0 = 0.07mm$) is imposed on the top of the block and the bottom of the block is only fix in the vertical direction. In the unloading process, the direction of vertical displacement boundary is changed to be positive. The block is discretized with 100 quadratic 9-node quadrilateral elements leading to 441 nodes.\\

\begin{figure}[!htb]
	\centering
	\def\svgwidth{0.45\columnwidth}
	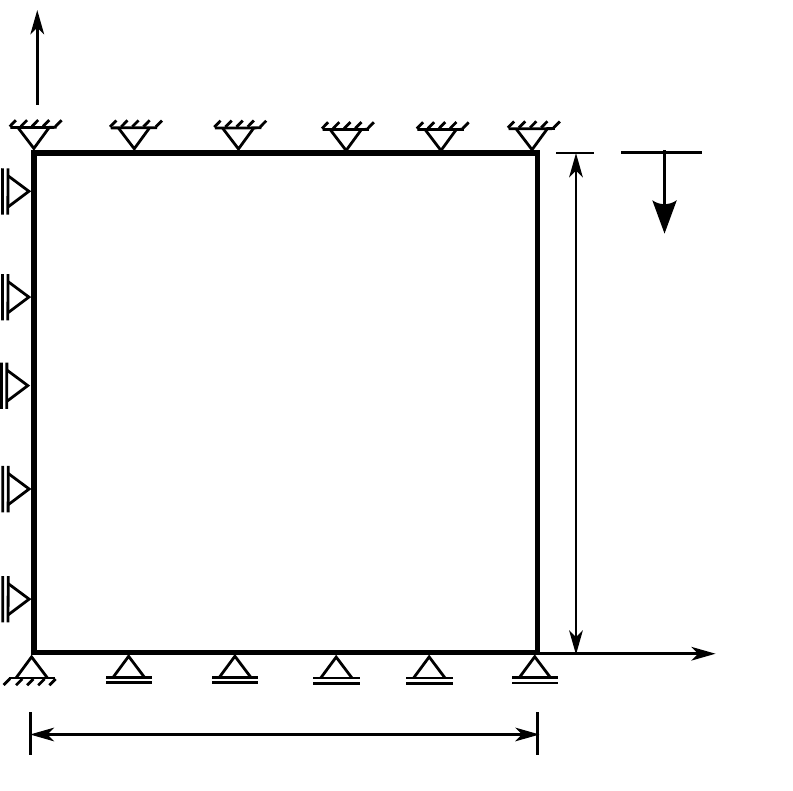
	\caption{2D punch problem.}
	\label{punch2Dgeo}
\end{figure}

Before unloading, the final deformation state of the block with ML based plasticity model is compared with that using the target plasticity model, which are depicted in Fig. \ref{punch2D}. It can be observed that the horizontal displacement of the structure is very close for the two models.\\

\begin{figure}[!htb]
	\centering
	\subfigure[With plasticity model]{\includegraphics[width=70mm]{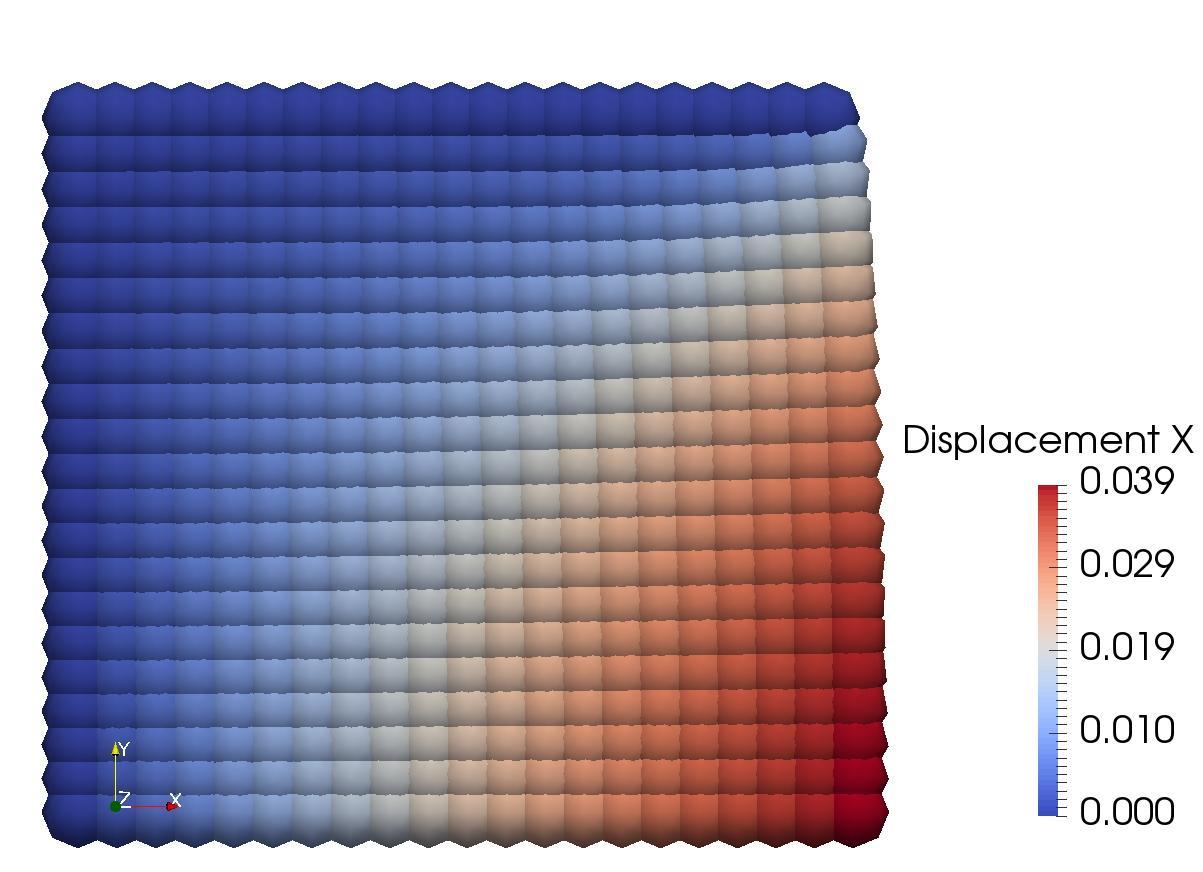}}
	\qquad
	\subfigure[With NN model]{\includegraphics[width=70mm]{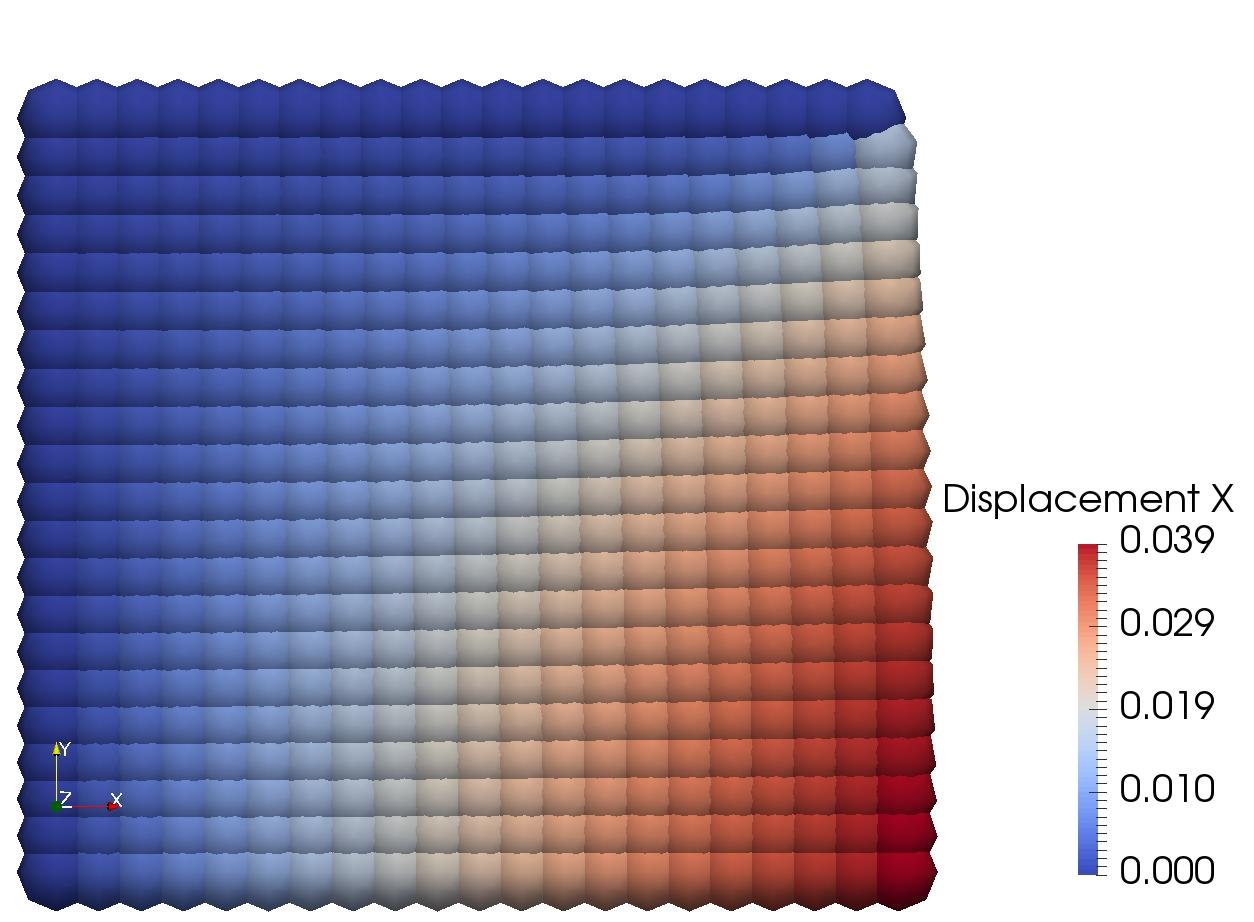}}
	\caption{Final deformation state of the 2D block.}
	\label{punch2D}
\end{figure}

The load displacement curve of the upper node $(0, 1)$ at the left end of the block is plotted in Fig. \ref{punch2dUL}, where it can be seen that the ML based model follows the plasticity model well both in loading and unloading. 

\begin{figure}[!htb]
	\centering
	\def\svgwidth{0.6\columnwidth}
	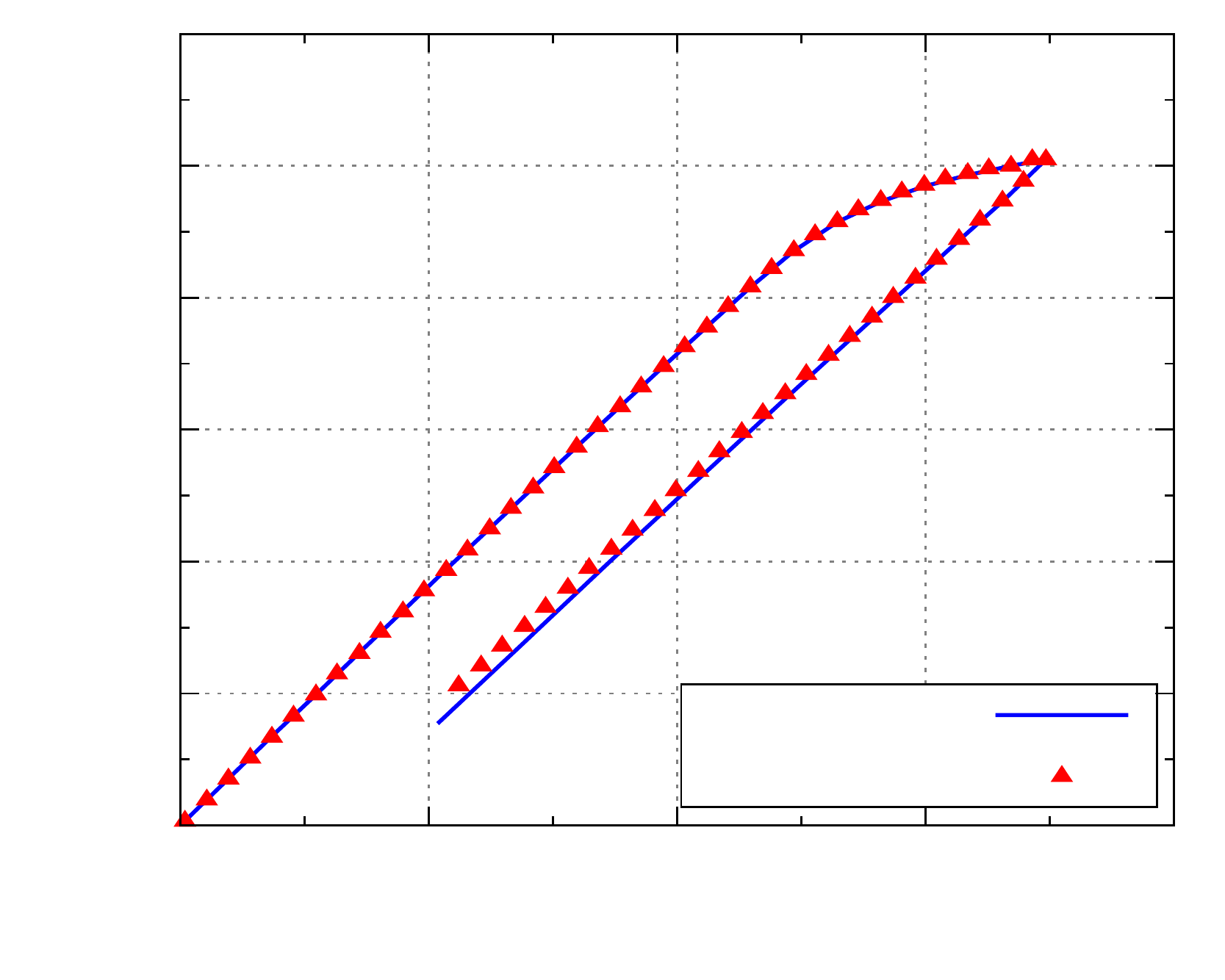
	\caption{Load deflection curve of the 2D block.}
	\label{punch2dUL}
\end{figure}

	\subsubsection{Applications in 3D finite element analysis}
	In this section, the ML based plasticity model is extended to 3D applications. To generate the training data, one hexahedron finite element is applied to different loading situations as described in Fig. \ref{ele8}. The von Mises plasticity with an exponential isotropic hardening law $\sigma_y = y_0+y_0(0.00002+\gamma)^{0.1}$ is set as the target model. The material parameters are set as: Young's modulus $E=10N/mm^2$, Poison's ratio $\nu=0.33$, and the initial yield stress $y_0=0.3MPa$. During the training data preparation, strain-stress sequences along 8100 loading paths are generated based on the sphere ($r=0.02$) in Fig. \ref{path-sphere}, where 90 values are assigned to the angles $\phi$ and $\theta$ respectively. Since the huge amount of data have to be collected for unloading in 3D, only the loading data is collected here and the unloading is not considered in this part.
	
	In the three-dimensional case, three FNNs are required to predict the coefficients, which are corresponding to the principle Cauchy stress components. The same network architecture (6-16-16-1) is employed for all FNNs, where three total strain components together with three accumulated absolute strains are applied as the input of the networks. The output of the neural network is the coefficient. 
	
	The weights are initialized by the Nguyen-Widrow method. During training, the Levenberg-Marquardt algorithm is applied as the optimizer, where the training progress is terminated when the gradient of global error is less than $10^{-7}$. The mean squared errors are decreased to $5.70\times 10^{-8}$, $1.13 \times 10^{-9}$ and $6.78 \times 10^{-10}$ for the first, second and third FNN respectively. The training process costs time of $1h20m21s$, $1h18m46s$ and $52m47s$ for the first, second and third FNN respectively.
	
	To evaluate the performance of the POD representation, the performances of training strain-stress model with one FNN(6-16-16-3) and training three strain-coefficient models with three FNNs(6-16-16-1) are compared. Without POD, only one FNN is required to approximate the mapping, where the output includes 3 stress components. With POD, three independent FNNs will be applied, where the output of FNN includes only one POD coefficient. The training performances within 2000 epochs are shown in Fig. \ref{3vs1}. It can be seen that the average of the mean squared errors of the three FNNs is smaller than that without POD. Additionally, training a FNN with architecture of (6-16-16-3) costs computation time of $4h22m43s$ whereas the average training time of the three FNNs (6-16-16-1) is $1h26m27$. It can be observed that the POD approach leads to less training time and better training performance.\\
	
	\begin{figure}[!htb]
		\centering
		\includegraphics[width=0.6\linewidth]{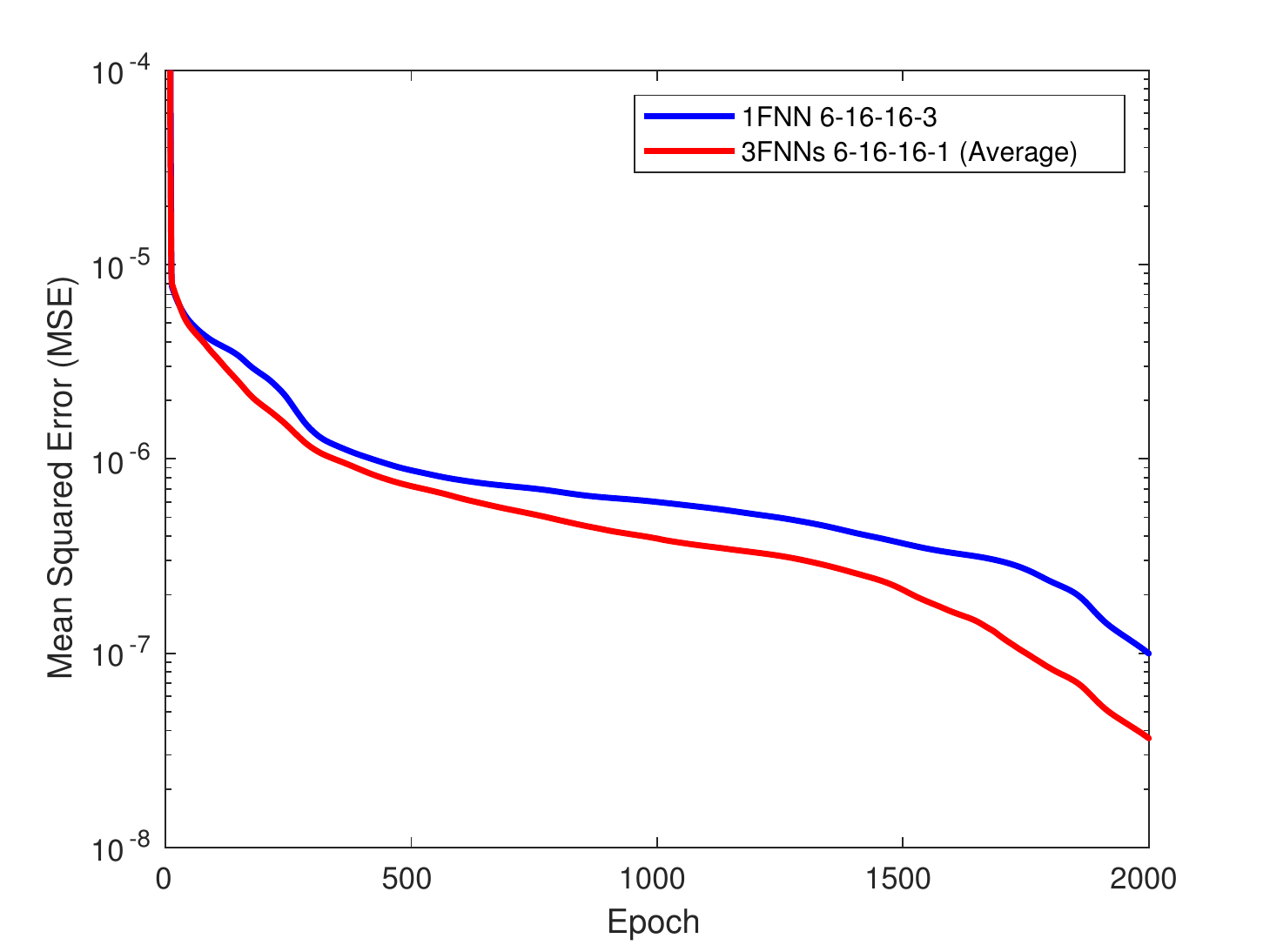}
		\caption{MSE of training FNN(6-16-16-3) and the average MSE of training 3 FNNs(6-16-16-1).}
		\label{3vs1}
	\end{figure}

	The first example to test the 3D machine learning based plasticity model is the necking of a bar as shown in Fig. \ref{bar3Dgeo}, where the left end of the bar is fixed and the displacement boundary $u_0=0.05mm$ is imposed at the right end along its axial direction. An artificial imperfection is set in the center of the bar to trigger the necking, where the radius at the center is chosen to be $R_c= 0.98R$. The bar is discretized with 200 quadratic 27-node elements leading to 2193 nodes.\\
	
	\begin{figure}[!htb]
		\centering
		\def\svgwidth{0.7\columnwidth}
		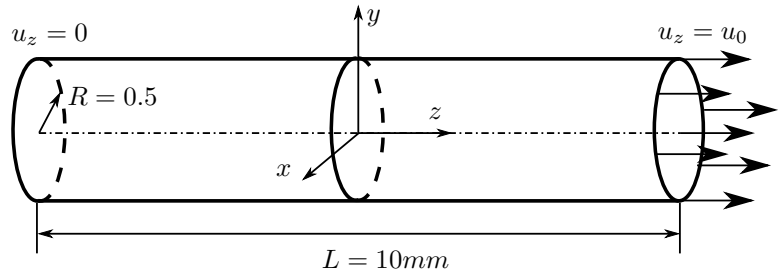
		\caption{Geometry and boundary conditions of the bar.}
		\label{bar3Dgeo}
	\end{figure}
	
	Fig. \ref{bar3D} shows the final deformation of the bar after tension, where only one quarter of the bar is computed due to the symmetry. It can be observed that the amounts of the necking computed by the two models are close to each other.\\
	
	\begin{figure}[!htb]
		\centering
		\subfigure[With plasticity model]{\label{plas}\includegraphics[width=120mm]{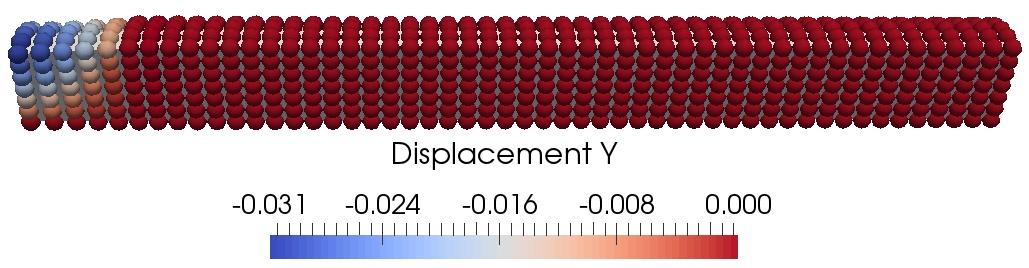}}
		\qquad\\
		\subfigure[With NN model]{\label{nn}\includegraphics[width=120mm]{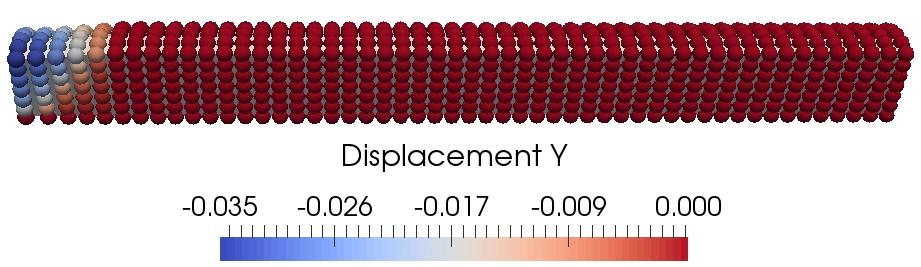}}
		\caption{Final deformation state of the cylindrical bar.}
		\label{bar3D}
	\end{figure}
	
	The load displacement curve of the bar under the uniaxial tension is plotted in Fig. \ref{bar3D-uf}, where the neural network based model follows the plasticity model quite well. 
	\begin{figure}[!htb]
		\centering
		\def\svgwidth{0.7\columnwidth}
		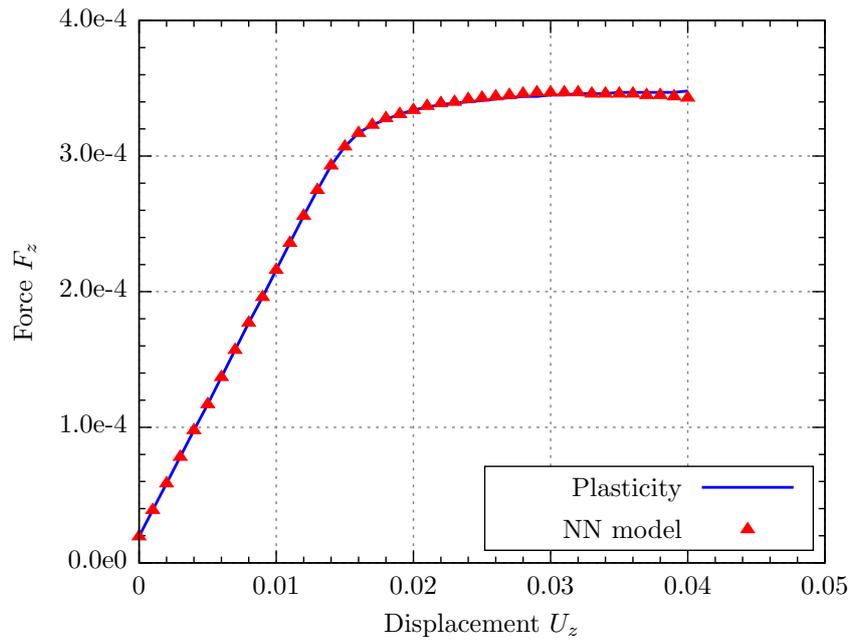
		\caption{Load deflection curve of the cylindrical bar.}
		\label{bar3D-uf}
	\end{figure}
	
	The second example is the punch test, where the vertical displacement boundary condition $(u_0=0.15mm)$ is imposed on the top of the block and the bottom of the block is only fixed in the vertical direction, as shown in Fig. \ref{punchgeo3D}. The block is discretized with 100 quadratic 27-node elements leading to 441 nodes.\\
	
	\begin{figure}[!htb]
		\centering
		\def\svgwidth{0.5\columnwidth}
		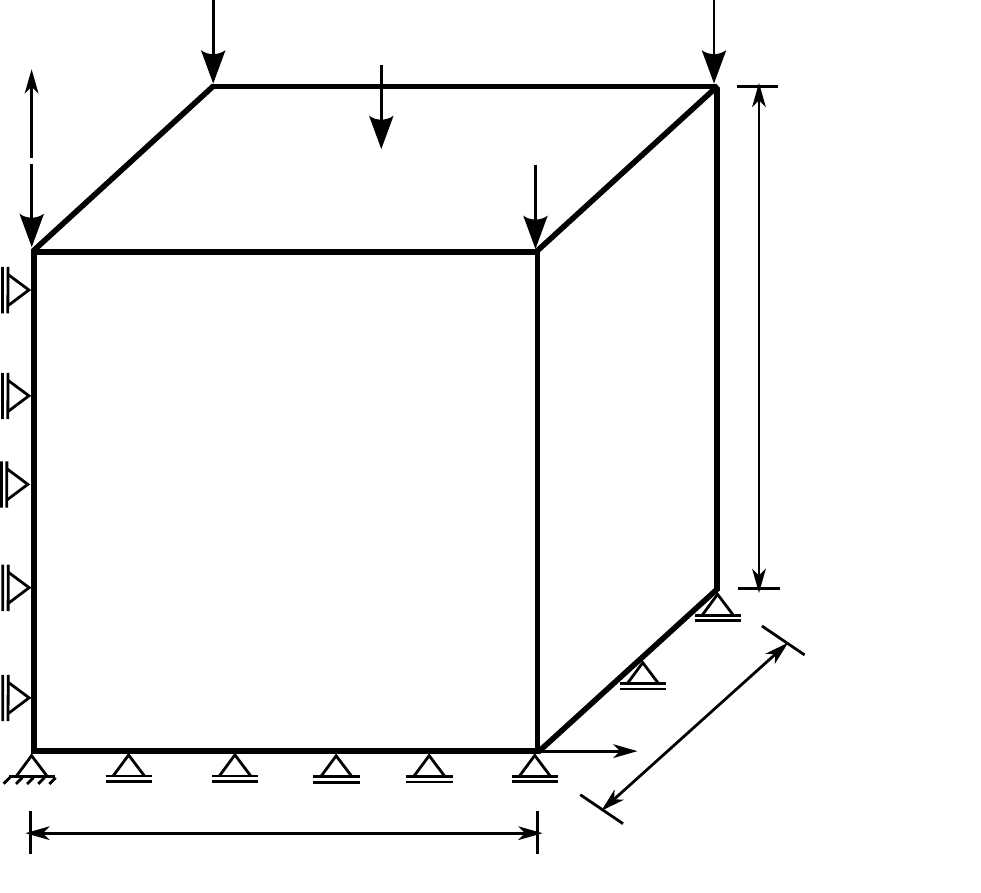
		\caption{3D punch problem.}
		\label{punchgeo3D}
	\end{figure}
	
	Fig. \ref{punch3D} shows the final deformation of the block after compression. It can be observed that the displacements in the horizontal direction computed by the two models are close to each other. The load displacement curve of the block under compression is plotted in Fig. \ref{punch3D-uf}. It can be seen that the neural network based model follows the plasticity model quite well. 
	
	\begin{figure}[!htb]
		\centering
		\subfigure[With plasticity model]{\label{punch3D}\includegraphics[width=75mm]{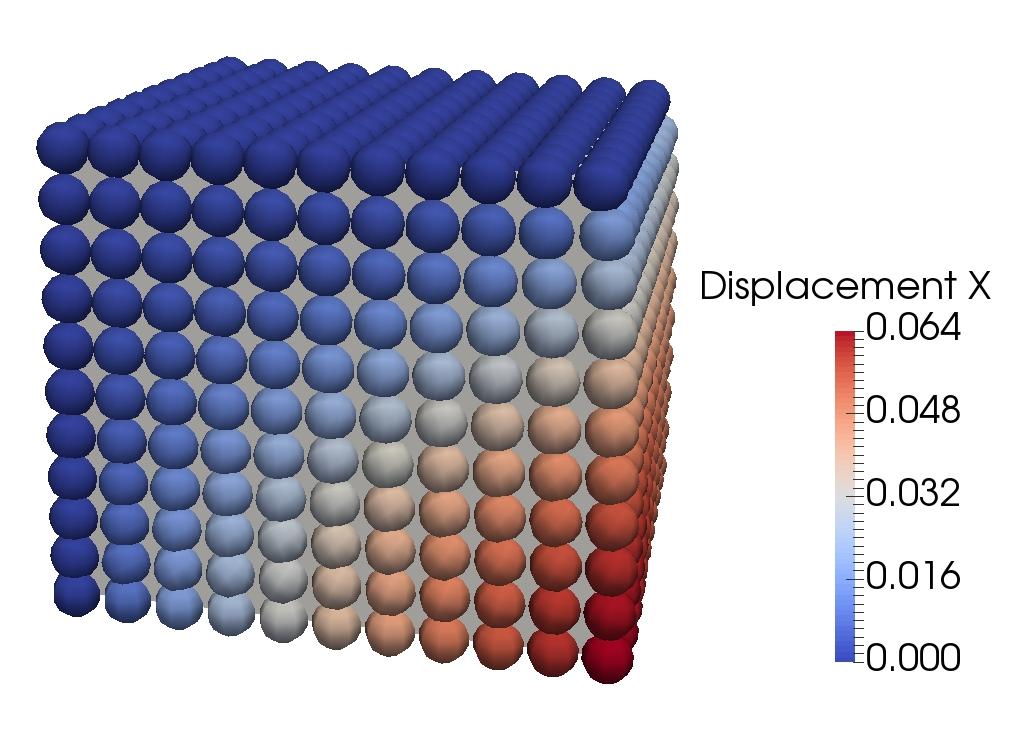}}
		\subfigure[With NN model]{\label{barnn}\includegraphics[width=75mm]{Fig/punch3D.jpg}}
		\caption{Final deformation state of the 3D punch problem.}
		\label{punch3D}
	\end{figure}
	
	\begin{figure}[!htb]
		\centering
		\def\svgwidth{0.7\columnwidth}
		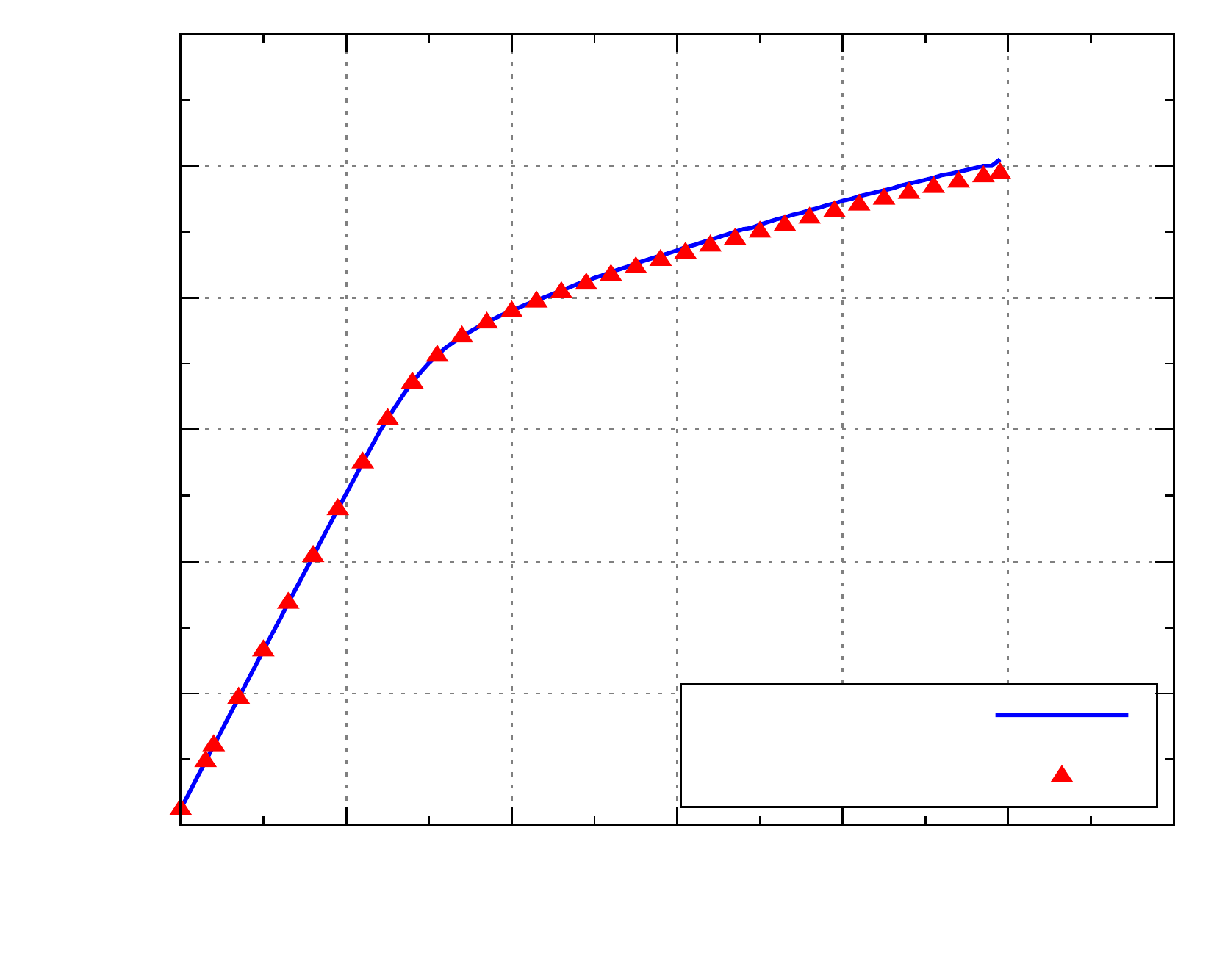
		\caption{Load deflection curve of the 3D punch problem.}
		\label{punch3D-uf}
	\end{figure}
	
	The last example for the 3D neural network based model is the Cook's membrane problem. The 3D beam is clamped at the left end and loaded at the right end by a constant distributed vertical load $q_0=0.03MPa$, as depicted in Fig. \ref{cookgeo3D}. By use of the quadratic finite element with 27 nodes, the beam is discretized using 1080 elements leading to 10309 nodes.\\
	
	\begin{figure}[!htb]
		\centering
		\def\svgwidth{0.6\columnwidth}
		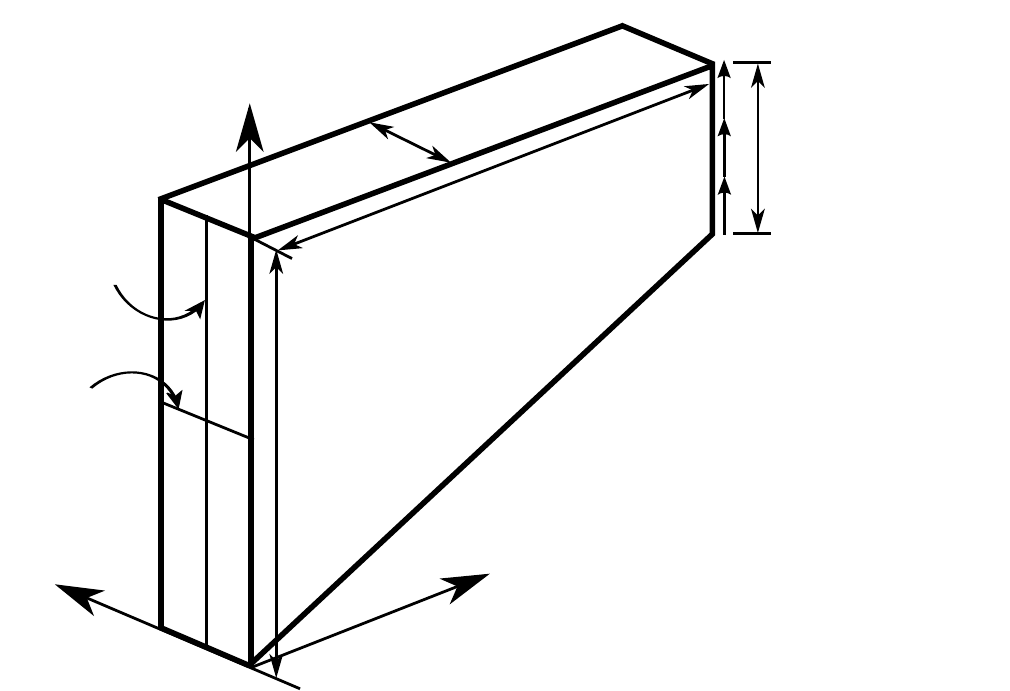
		\caption{3D Cook's membrane problem.}
		\label{cookgeo3D}
	\end{figure}
	Fig. \ref{cook3D} shows the final deformation of the Cook's membrane. It can be observed that the displacements in the vertical direction computed by the two models are almost identical. The load displacement curve of the upper node (48,0,60) is plotted in Fig. \ref{cook3D-uf}. It can be seen that the machine learning based model follows the plasticity model quite well. 
	
	\begin{figure}[!htb]
		\centering
		\subfigure[With plasticity model]{\label{a}\includegraphics[width=70mm]{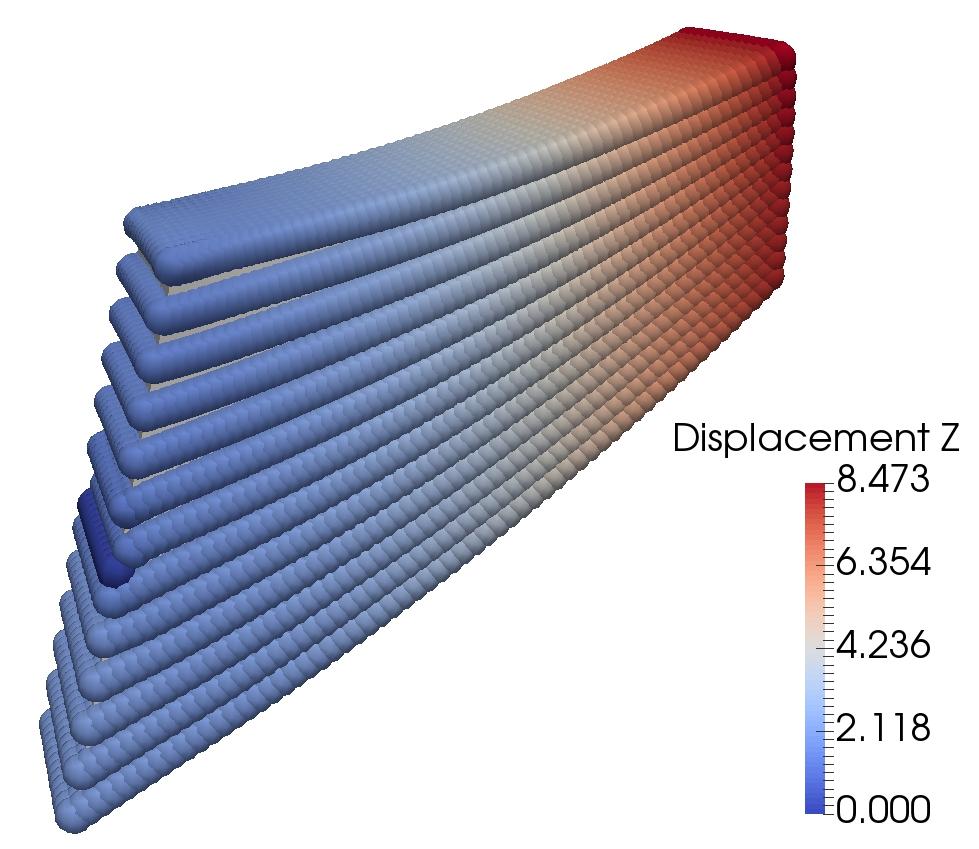}}
		\qquad 
		\subfigure[With NN model]{\label{cook3DNN}\includegraphics[width=70mm]{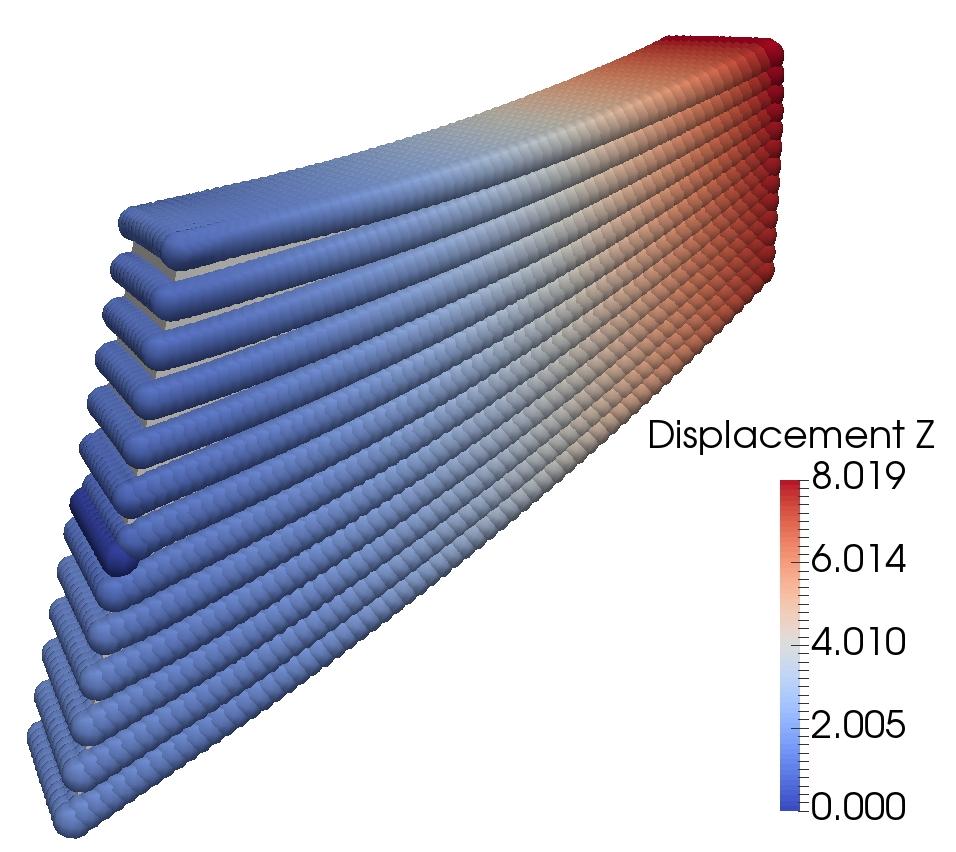}}
		\caption{Final deformation state of the 3D Cook's membrane.}
		\label{cook3D}
	\end{figure}
	
	\begin{figure}[!htb]
		\centering
		\def\svgwidth{0.6\columnwidth}
		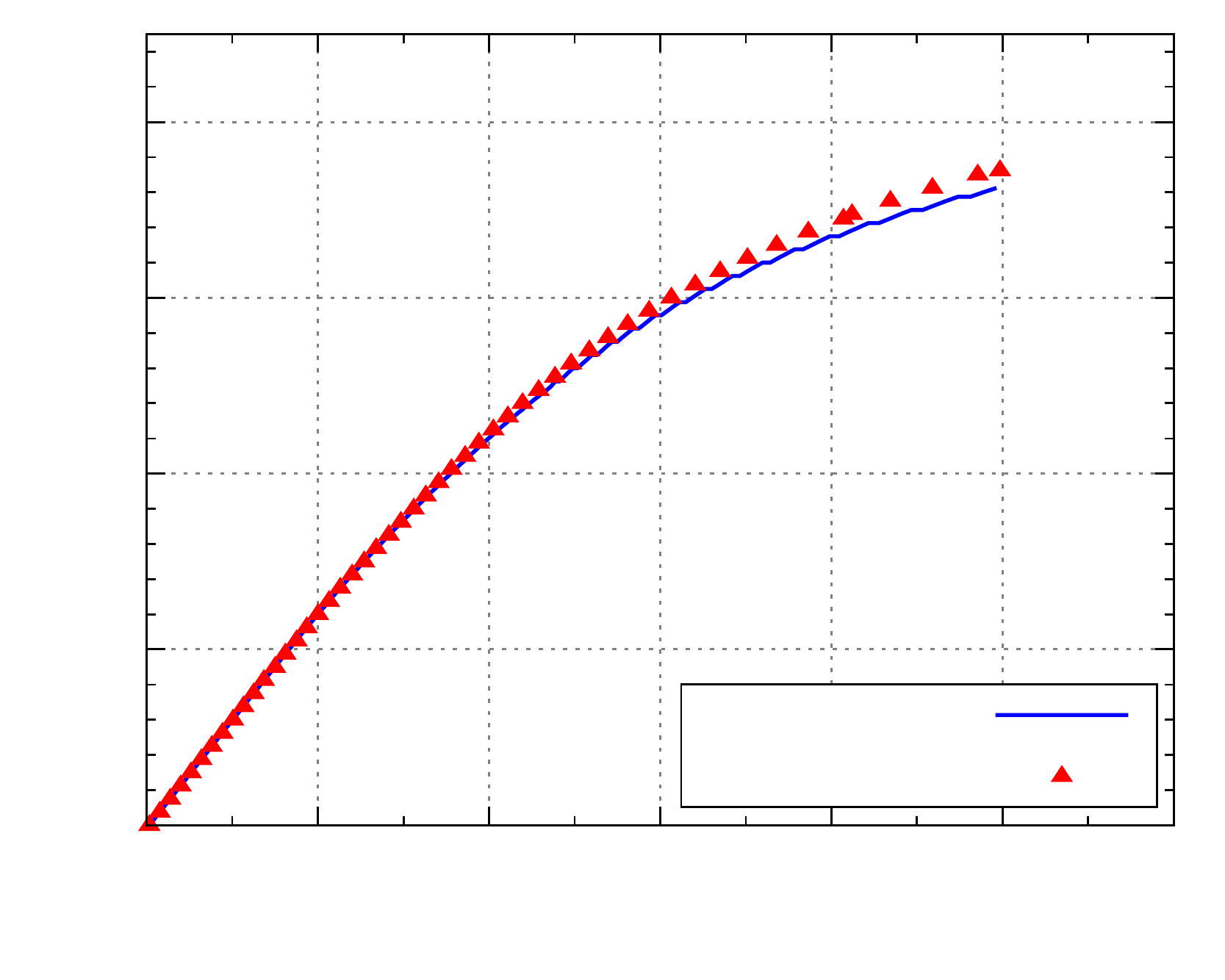
		\caption{Load deflection curve of the 3D Cook's membrane.}
		\label{cook3D-uf}
	\end{figure}
	
	\section{Conclusions}
	In this work, a machine learning based material modelling approach for hyper-elasticity and plasticity is proposed. Common tools such as FNNs show proficient performances for capturing the mapping between strain and stress in the case of elasticity. However, history variables are required to distinguish the loading history in case of plasticity. In this context, FNNs show subpar performances. Thus, the accumulated absolute strain is proposed to be the history variable, which captures the loading history well without requirement for additional data. Here we present a novel method called Proper Orthogonal Decomposition Feed forward Neural Network (PODFNN), which in combination with the introduced history variable is able to overcome this problem. By use of the POD, less training time and better training performance are obtained in the network training. Additionally, it has been shown that the training data collected only from the multi-axial loading tests are enough to capture the von Mises yield surface and the hardening law. The automatic symbolic differentiation tool AceGen provides a very convenient way to derive the tangent matrix for the machine learning based material model. The generalization and accuracy of the presented model as well as the data generation strategy have been verified by finite element applications both in 2D and 3D.
		
	\section*{Acknowledgements}
	The first author would like to thank the China Scholarship Council (CSC) and the Graduate Academy of Leibniz Universit\"at Hannover for the financial support. The second author acknowledges the financial support from the Deutsche Forschungsgemeinschaft under Germanys Excellence Strategy within the Cluster of Excellence PhoenixD (EXC 2122, Project ID 390833453). The last author acknowledges the support of Deutsche Forschungsgemeinschaft for the project C2 within the collaborative research center/Transregio TR73.

\clearpage
\small

\bibliographystyle{plainnat}
\bibliography{literature}

\end{document}

%% file: definitions.tex
\long\def\symbolfootnote[#1]#2{\begingroup \def\thefootnote{\fnsymbol{footnote}}\footnote[#1]{#2} \endgroup}

%% file: Fig/flowchart.pdf_tex
\begingroup%
  \makeatletter%
  \providecommand\color[2][]{%
    \errmessage{(Inkscape) Color is used for the text in Inkscape, but the package 'color.sty' is not loaded}%
    \renewcommand\color[2][]{}%
  }%
  \providecommand\transparent[1]{%
    \errmessage{(Inkscape) Transparency is used (non-zero) for the text in Inkscape, but the package 'transparent.sty' is not loaded}%
    \renewcommand\transparent[1]{}%
  }%
  \providecommand\rotatebox[2]{#2}%
  \ifx\svgwidth\undefined%
    \setlength{\unitlength}{310.1838623bp}%
    \ifx\svgscale\undefined%
      \relax%
    \else%
      \setlength{\unitlength}{\unitlength * \real{\svgscale}}%
    \fi%
  \else%
    \setlength{\unitlength}{\svgwidth}%
  \fi%
  \global\let\svgwidth\undefined%
  \global\let\svgscale\undefined%
  \makeatother%
  \begin{picture}(1,1.02844941)%
    \put(0,0){\includegraphics[width=\unitlength]{flowchart.pdf}}%
    \put(0.01878237,0.98542447){\color[rgb]{0,0,0}\makebox(0,0)[lb]{\smash{\textcolor{white}{\textbf{Data Collection}}}}}%
    \put(0.05393022,0.84781751){\color[rgb]{0,0,0}\makebox(0,0)[lb]{\smash{\begin{minipage}{50mm}
Loading conditions\\
Data from experiments\\
Data from simulations\\
Data preprocessing
\end{minipage}}}}%
    \put(0.74471794,0.75822174){\color[rgb]{0,0,0}\makebox(0,0)[lb]{\smash{$\varepsilon$}}}%
    \put(0.52277777,0.92995066){\color[rgb]{0,0,0}\makebox(0,0)[lb]{\smash{$\sigma$}}}%
    \put(0.02110549,0.6312066){\color[rgb]{0,0,0}\makebox(0,0)[lb]{\smash{\textcolor{white}{\textbf{Machine Learning}}}}}%
    \put(0.04653675,0.5042412){\color[rgb]{0,0,0}\makebox(0,0)[lb]{\smash{\begin{minipage}{50mm}
Artificial neural networks\\
Network architecture\\
Training algorithm ...
\end{minipage}}}}%
    \put(0.74640805,0.52773305){\color[rgb]{0,0,0}\makebox(0,0)[lb]{\smash{$\sigma_1$}}}%
    \put(0.74739741,0.48283733){\color[rgb]{0,0,0}\makebox(0,0)[lb]{\smash{$\sigma_2$}}}%
    \put(0.74814726,0.43577917){\color[rgb]{0,0,0}\makebox(0,0)[lb]{\smash{$\sigma_3$}}}%
    \put(0.55525095,0.4339566){\color[rgb]{0,0,0}\makebox(0,0)[lb]{\smash{$\varepsilon_3$}}}%
    \put(0.55273936,0.48158232){\color[rgb]{0,0,0}\makebox(0,0)[lb]{\smash{$\varepsilon_2$}}}%
    \put(0.55450723,0.52768511){\color[rgb]{0,0,0}\makebox(0,0)[lb]{\smash{$\varepsilon_1$}}}%
    \put(0.02563413,0.27778781){\color[rgb]{0,0,0}\makebox(0,0)[lb]{\smash{\textcolor{white}{\textbf{Validation}}}}}%
    \put(0.04803473,0.13208416){\color[rgb]{0,0,0}\makebox(0,0)[lb]{\smash{\begin{minipage}{50mm}
Derive the material tangent\\
Incorporartion in FEM code\\
Benchmark tests\\
Model valitation
\end{minipage}}}}%
    \put(0.30096864,0.40575304){\color[rgb]{0,0,0}\makebox(0,0)[lb]{\smash{$\boldsymbol \sigma=f(\boldsymbol \varepsilon, \boldsymbol W)$}}}%
    \put(0.76138853,0.18746068){\color[rgb]{0,0,0}\makebox(0,0)[lb]{\smash{$F$}}}%
    \put(0.84102321,0.87737305){\color[rgb]{0,0,0}\makebox(0,0)[lb]{\smash{{\small Data supplement}}}}%
    \put(0.8286641,0.51353839){\color[rgb]{0,0,0}\makebox(0,0)[lb]{\smash{{\small Accuracy improvement}}}}%
    \put(0.05184968,0.40582081){\color[rgb]{0,0,0}\makebox(0,0)[lb]{\smash{{Material model:}}}}%
  \end{picture}%
\endgroup%

%% file: Fig/fnntrain.pdf_tex
\begingroup%
  \makeatletter%
  \providecommand\color[2][]{%
    \errmessage{(Inkscape) Color is used for the text in Inkscape, but the package 'color.sty' is not loaded}%
    \renewcommand\color[2][]{}%
  }%
  \providecommand\transparent[1]{%
    \errmessage{(Inkscape) Transparency is used (non-zero) for the text in Inkscape, but the package 'transparent.sty' is not loaded}%
    \renewcommand\transparent[1]{}%
  }%
  \providecommand\rotatebox[2]{#2}%
  \ifx\svgwidth\undefined%
    \setlength{\unitlength}{625.90352783bp}%
    \ifx\svgscale\undefined%
      \relax%
    \else%
      \setlength{\unitlength}{\unitlength * \real{\svgscale}}%
    \fi%
  \else%
    \setlength{\unitlength}{\svgwidth}%
  \fi%
  \global\let\svgwidth\undefined%
  \global\let\svgscale\undefined%
  \makeatother%
  \begin{picture}(1,0.34122784)%
    \put(0,0){\includegraphics[width=\unitlength]{fnntrain.pdf}}%
    \put(0.0216344,0.32180592){\color[rgb]{0,0,0}\makebox(0,0)[lb]{\smash{Input layer}}}%
    \put(0.40558225,0.31967135){\color[rgb]{0,0,0}\makebox(0,0)[lb]{\smash{Output layer}}}%
    \put(0.20534531,0.31852012){\color[rgb]{0,0,0}\makebox(0,0)[lb]{\smash{Hidden layers}}}%
    \put(0.58561086,0.3166664){\color[rgb]{0,0,0}\makebox(0,0)[lb]{\smash{Targets}}}%
    \put(0.07309657,0.03107356){\color[rgb]{0,0,0}\makebox(0,0)[lb]{\smash{b}}}%
    \put(0.21777851,0.00973628){\color[rgb]{0,0,0}\makebox(0,0)[lb]{\smash{b}}}%
    \put(0.35251155,0.00890514){\color[rgb]{0,0,0}\makebox(0,0)[lb]{\smash{b}}}%
    \put(0.00469665,0.25416749){\color[rgb]{0,0,0}\makebox(0,0)[lb]{\smash{$i_1$}}}%
    \put(0.00348641,0.17969854){\color[rgb]{0,0,0}\makebox(0,0)[lb]{\smash{$i_2$}}}%
    \put(-0.00022113,0.08555858){\color[rgb]{0,0,0}\makebox(0,0)[lb]{\smash{$i_m$}}}%
    \put(-0.00092366,0.02514042){\color[rgb]{0,0,0}\makebox(0,0)[lb]{\smash{bias}}}%
    \put(0.50637158,0.23596498){\color[rgb]{0,0,0}\makebox(0,0)[lb]{\smash{$o_1$}}}%
    \put(0.50351025,0.16802245){\color[rgb]{0,0,0}\makebox(0,0)[lb]{\smash{$o_2$}}}%
    \put(0.50606655,0.07727364){\color[rgb]{0,0,0}\makebox(0,0)[lb]{\smash{$o_n$}}}%
    \put(0.64282884,0.23768175){\color[rgb]{0,0,0}\makebox(0,0)[lb]{\smash{$t_1$}}}%
    \put(0.63963346,0.17249599){\color[rgb]{0,0,0}\makebox(0,0)[lb]{\smash{$t_2$}}}%
    \put(0.63963346,0.07471734){\color[rgb]{0,0,0}\makebox(0,0)[lb]{\smash{$t_n$}}}%
    \put(0.75274994,0.16227077){\color[rgb]{0,0,0}\makebox(0,0)[lb]{\smash{$E(\boldsymbol w)= \frac{1}{n} \sum\limits_{j=1}^{n} \left[ o_j(\boldsymbol w)-t_j \right]^2 $}}}%
  \end{picture}%
\endgroup%

%% file: Fig/neuron.pdf_tex
\begingroup%
  \makeatletter%
  \providecommand\color[2][]{%
    \errmessage{(Inkscape) Color is used for the text in Inkscape, but the package 'color.sty' is not loaded}%
    \renewcommand\color[2][]{}%
  }%
  \providecommand\transparent[1]{%
    \errmessage{(Inkscape) Transparency is used (non-zero) for the text in Inkscape, but the package 'transparent.sty' is not loaded}%
    \renewcommand\transparent[1]{}%
  }%
  \providecommand\rotatebox[2]{#2}%
  \ifx\svgwidth\undefined%
    \setlength{\unitlength}{332.52099609bp}%
    \ifx\svgscale\undefined%
      \relax%
    \else%
      \setlength{\unitlength}{\unitlength * \real{\svgscale}}%
    \fi%
  \else%
    \setlength{\unitlength}{\svgwidth}%
  \fi%
  \global\let\svgwidth\undefined%
  \global\let\svgscale\undefined%
  \makeatother%
  \begin{picture}(1,0.46893009)%
    \put(0,0){\includegraphics[width=\unitlength]{neuron.pdf}}%
    \put(0.36999675,0.22211987){\color[rgb]{0,0,0}\makebox(0,0)[lb]{\smash{$\sum$}}}%
    \put(0.49389268,0.22044367){\color[rgb]{0,0,0}\makebox(0,0)[lb]{\smash{$f(x)$}}}%
    \put(0.18639911,0.37069235){\color[rgb]{0,0,0}\makebox(0,0)[lb]{\smash{$w_{1j}$}}}%
    \put(0.11790983,0.31469744){\color[rgb]{0,0,0}\makebox(0,0)[lb]{\smash{$w_{2j}$}}}%
    \put(0.13492195,0.21291727){\color[rgb]{0,0,0}\makebox(0,0)[lb]{\smash{$w_{3j}$}}}%
    \put(0.80689659,0.21972209){\color[rgb]{0,0,0}\makebox(0,0)[lb]{\smash{$o_j^k$}}}%
    \put(0.00444014,0.43237225){\color[rgb]{0,0,0}\makebox(0,0)[lb]{\smash{$o_1^{k-1}$}}}%
    \put(-0.00159764,0.30818469){\color[rgb]{0,0,0}\makebox(0,0)[lb]{\smash{$o_2^{k-1}$}}}%
    \put(0.01031075,0.16698473){\color[rgb]{0,0,0}\makebox(0,0)[lb]{\smash{$o_3^{k-1}$}}}%
    \put(0.03072512,0.00707192){\color[rgb]{0,0,0}\makebox(0,0)[lb]{\smash{$...$}}}%
    \put(0.17769378,0.06277461){\color[rgb]{0,0,0}\makebox(0,0)[lb]{\smash{$...$}}}%
  \end{picture}%
\endgroup%

%% file: Fig/sequence1d.pdf_tex
\begingroup%
  \makeatletter%
  \providecommand\color[2][]{%
    \errmessage{(Inkscape) Color is used for the text in Inkscape, but the package 'color.sty' is not loaded}%
    \renewcommand\color[2][]{}%
  }%
  \providecommand\transparent[1]{%
    \errmessage{(Inkscape) Transparency is used (non-zero) for the text in Inkscape, but the package 'transparent.sty' is not loaded}%
    \renewcommand\transparent[1]{}%
  }%
  \providecommand\rotatebox[2]{#2}%
  \ifx\svgwidth\undefined%
    \setlength{\unitlength}{141.56276855bp}%
    \ifx\svgscale\undefined%
      \relax%
    \else%
      \setlength{\unitlength}{\unitlength * \real{\svgscale}}%
    \fi%
  \else%
    \setlength{\unitlength}{\svgwidth}%
  \fi%
  \global\let\svgwidth\undefined%
  \global\let\svgscale\undefined%
  \makeatother%
  \begin{picture}(1,0.93273643)%
    \put(0,0){\includegraphics[width=\unitlength]{sequence1d.pdf}}%
    \put(0.16183072,0.07887333){\color[rgb]{0,0,0}\makebox(0,0)[lb]{\smash{$\varepsilon^1$}}}%
    \put(0.83833105,0.16339459){\color[rgb]{0,0,0}\makebox(0,0)[lb]{\smash{$\varepsilon$}}}%
    \put(0.04780655,0.07443781){\color[rgb]{0,0,0}\makebox(0,0)[lb]{\smash{$O$}}}%
    \put(0.29560995,0.08065003){\color[rgb]{0,0,0}\makebox(0,0)[lb]{\smash{$\varepsilon^2$}}}%
    \put(0.43383541,0.0790965){\color[rgb]{0,0,0}\makebox(0,0)[lb]{\smash{$\varepsilon^3$}}}%
    \put(0.59492627,0.08065003){\color[rgb]{0,0,0}\makebox(0,0)[lb]{\smash{$\varepsilon^t$}}}%
    \put(0.05022135,0.41810213){\color[rgb]{0,0,0}\makebox(0,0)[lb]{\smash{$\sigma^1$}}}%
    \put(0.04901136,0.60913257){\color[rgb]{0,0,0}\makebox(0,0)[lb]{\smash{$\sigma^2$}}}%
    \put(0.04901153,0.70231827){\color[rgb]{0,0,0}\makebox(0,0)[lb]{\smash{$\sigma^3$}}}%
    \put(0.04901136,0.77220725){\color[rgb]{0,0,0}\makebox(0,0)[lb]{\smash{$\sigma^t$}}}%
    \put(0.14884009,0.8731587){\color[rgb]{0,0,0}\makebox(0,0)[lb]{\smash{$\sigma$}}}%
  \end{picture}%
\endgroup%

%% file: Fig/bargeo.pdf_tex
\begingroup%
  \makeatletter%
  \providecommand\color[2][]{%
    \errmessage{(Inkscape) Color is used for the text in Inkscape, but the package 'color.sty' is not loaded}%
    \renewcommand\color[2][]{}%
  }%
  \providecommand\transparent[1]{%
    \errmessage{(Inkscape) Transparency is used (non-zero) for the text in Inkscape, but the package 'transparent.sty' is not loaded}%
    \renewcommand\transparent[1]{}%
  }%
  \providecommand\rotatebox[2]{#2}%
  \ifx\svgwidth\undefined%
    \setlength{\unitlength}{221.06220703bp}%
    \ifx\svgscale\undefined%
      \relax%
    \else%
      \setlength{\unitlength}{\unitlength * \real{\svgscale}}%
    \fi%
  \else%
    \setlength{\unitlength}{\svgwidth}%
  \fi%
  \global\let\svgwidth\undefined%
  \global\let\svgscale\undefined%
  \makeatother%
  \begin{picture}(1,1.61769883)%
    \put(0,0){\includegraphics[width=\unitlength]{bargeo.pdf}}%
    \put(0.26515488,0.01063756){\color[rgb]{0,0,0}\makebox(0,0)[lb]{\smash{$W=4mm$}}}%
    \put(0.78924555,0.82872137){\color[rgb]{0,0,0}\makebox(0,0)[lb]{\smash{$L=10mm$}}}%
    \put(0.35814032,1.56270865){\color[rgb]{0,0,0}\makebox(0,0)[lb]{\smash{$q_0$}}}%
  \end{picture}%
\endgroup%

%% file: Fig/cookgeo.pdf_tex
\begingroup%
  \makeatletter%
  \providecommand\color[2][]{%
    \errmessage{(Inkscape) Color is used for the text in Inkscape, but the package 'color.sty' is not loaded}%
    \renewcommand\color[2][]{}%
  }%
  \providecommand\transparent[1]{%
    \errmessage{(Inkscape) Transparency is used (non-zero) for the text in Inkscape, but the package 'transparent.sty' is not loaded}%
    \renewcommand\transparent[1]{}%
  }%
  \providecommand\rotatebox[2]{#2}%
  \ifx\svgwidth\undefined%
    \setlength{\unitlength}{195.0685791bp}%
    \ifx\svgscale\undefined%
      \relax%
    \else%
      \setlength{\unitlength}{\unitlength * \real{\svgscale}}%
    \fi%
  \else%
    \setlength{\unitlength}{\svgwidth}%
  \fi%
  \global\let\svgwidth\undefined%
  \global\let\svgscale\undefined%
  \makeatother%
  \begin{picture}(1,1.19470241)%
    \put(0,0){\includegraphics[width=\unitlength]{cookgeo.pdf}}%
    \put(0.33821213,0.21718599){\color[rgb]{0,0,0}\makebox(0,0)[lb]{\smash{$x$}}}%
    \put(0.09025178,1.05574237){\color[rgb]{0,0,0}\makebox(0,0)[lb]{\smash{$y$}}}%
    \put(0.29312833,0.01205506){\color[rgb]{0,0,0}\makebox(0,0)[lb]{\smash{$48mm$}}}%
    \put(0.81672273,0.42874395){\color[rgb]{0,0,0}\makebox(0,0)[lb]{\smash{$44mm$}}}%
    \put(0.81835315,0.91598295){\color[rgb]{0,0,0}\makebox(0,0)[lb]{\smash{$16mm$}}}%
    \put(0.71466077,1.13238458){\color[rgb]{0,0,0}\makebox(0,0)[lb]{\smash{$q_0$}}}%
  \end{picture}%
\endgroup%

%% file: Fig/sequence1dplas.pdf_tex
\begingroup%
  \makeatletter%
  \providecommand\color[2][]{%
    \errmessage{(Inkscape) Color is used for the text in Inkscape, but the package 'color.sty' is not loaded}%
    \renewcommand\color[2][]{}%
  }%
  \providecommand\transparent[1]{%
    \errmessage{(Inkscape) Transparency is used (non-zero) for the text in Inkscape, but the package 'transparent.sty' is not loaded}%
    \renewcommand\transparent[1]{}%
  }%
  \providecommand\rotatebox[2]{#2}%
  \ifx\svgwidth\undefined%
    \setlength{\unitlength}{140.22490234bp}%
    \ifx\svgscale\undefined%
      \relax%
    \else%
      \setlength{\unitlength}{\unitlength * \real{\svgscale}}%
    \fi%
  \else%
    \setlength{\unitlength}{\svgwidth}%
  \fi%
  \global\let\svgwidth\undefined%
  \global\let\svgscale\undefined%
  \makeatother%
  \begin{picture}(1,1.09727836)%
    \put(0,0){\includegraphics[width=\unitlength]{sequence1dplas.pdf}}%
    \put(0.11433467,0.23997251){\color[rgb]{0,0,0}\makebox(0,0)[lb]{\smash{$\varepsilon^1$}}}%
    \put(0.84632945,0.32059634){\color[rgb]{0,0,0}\makebox(0,0)[lb]{\smash{$\varepsilon$}}}%
    \put(0.04826266,0.23079083){\color[rgb]{0,0,0}\makebox(0,0)[lb]{\smash{$O$}}}%
    \put(0.20558056,0.24098198){\color[rgb]{0,0,0}\makebox(0,0)[lb]{\smash{$\varepsilon^2$}}}%
    \put(0.36872912,0.2394138){\color[rgb]{0,0,0}\makebox(0,0)[lb]{\smash{$\varepsilon^3$}}}%
    \put(0.54649304,0.23945629){\color[rgb]{0,0,0}\makebox(0,0)[lb]{\smash{$\varepsilon^4$}}}%
    \put(0.04721558,0.28958746){\color[rgb]{0,0,0}\makebox(0,0)[lb]{\smash{$\sigma^1$}}}%
    \put(0.04947897,0.67930511){\color[rgb]{0,0,0}\makebox(0,0)[lb]{\smash{$\sigma^2$}}}%
    \put(0.05026314,0.84663083){\color[rgb]{0,0,0}\makebox(0,0)[lb]{\smash{$\sigma^3$}}}%
    \put(0.04947897,0.91239711){\color[rgb]{0,0,0}\makebox(0,0)[lb]{\smash{$\sigma^4$}}}%
    \put(0.15026015,1.03713221){\color[rgb]{0,0,0}\makebox(0,0)[lb]{\smash{$\sigma$}}}%
    \put(0.04022556,0.40861674){\color[rgb]{0,0,0}\makebox(0,0)[lb]{\smash{$\sigma^6$}}}%
    \put(0.05228731,0.96172026){\color[rgb]{0,0,0}\makebox(0,0)[lb]{\smash{$\sigma^5$}}}%
    \put(0.6714059,0.24247111){\color[rgb]{0,0,0}\makebox(0,0)[lb]{\smash{$\varepsilon^5$}}}%
    \put(0.53446873,0.19414525){\color[rgb]{0,0,0}\makebox(0,0)[lb]{\smash{$(\varepsilon^6)$}}}%
  \end{picture}%
\endgroup%

%% file: Fig/von_mises_yieldsurf.pdf_tex
\begingroup%
  \makeatletter%
  \providecommand\color[2][]{%
    \errmessage{(Inkscape) Color is used for the text in Inkscape, but the package 'color.sty' is not loaded}%
    \renewcommand\color[2][]{}%
  }%
  \providecommand\transparent[1]{%
    \errmessage{(Inkscape) Transparency is used (non-zero) for the text in Inkscape, but the package 'transparent.sty' is not loaded}%
    \renewcommand\transparent[1]{}%
  }%
  \providecommand\rotatebox[2]{#2}%
  \ifx\svgwidth\undefined%
    \setlength{\unitlength}{607.30775452bp}%
    \ifx\svgscale\undefined%
      \relax%
    \else%
      \setlength{\unitlength}{\unitlength * \real{\svgscale}}%
    \fi%
  \else%
    \setlength{\unitlength}{\svgwidth}%
  \fi%
  \global\let\svgwidth\undefined%
  \global\let\svgscale\undefined%
  \makeatother%
  \begin{picture}(1,0.68432513)%
    \put(0,0){\includegraphics[width=\unitlength]{von_mises_yieldsurf.pdf}}%
    \put(0.72820617,0.55269689){\color[rgb]{0,0,0}\makebox(0,0)[lb]{\smash{Equal biaxial tension}}}%
    \put(0.62888208,0.19226313){\color[rgb]{0,0,0}\makebox(0,0)[lb]{\smash{Pure shear}}}%
    \put(0.07696784,0.09293898){\color[rgb]{0,0,0}\makebox(0,0)[lt]{\begin{minipage}{0.34712233\unitlength}\raggedright Equal biaxial compression\end{minipage}}}%
    \put(0.72615825,0.31206639){\color[rgb]{0,0,0}\makebox(0,0)[lt]{\begin{minipage}{0.21400757\unitlength}\raggedright Uniaxial Tension\end{minipage}}}%
    \put(0.51727048,0.11853803){\color[rgb]{0,0,0}\makebox(0,0)[lb]{\smash{Uniaxial compression}}}%
    \put(-0.00228982,0.34780445){\color[rgb]{0,0,0}\makebox(0,0)[lb]{\smash{Uniaxial compression}}}%
    \put(0.21932849,0.43843125){\color[rgb]{0,0,0}\makebox(0,0)[lb]{\smash{Pure shear}}}%
    \put(0.26829895,0.52095415){\color[rgb]{0,0,0}\makebox(0,0)[lb]{\smash{Uniaxial tension}}}%
    \put(0.51522259,0.66430851){\color[rgb]{0,0,0}\makebox(0,0)[lb]{\smash{$\sigma_2$}}}%
    \put(0.83572195,0.33868932){\color[rgb]{0,0,0}\makebox(0,0)[lb]{\smash{$\sigma_1$}}}%
    \put(0.51010277,0.34380913){\color[rgb]{0,0,0}\makebox(0,0)[lb]{\smash{$O$}}}%
  \end{picture}%
\endgroup%

%% file: Fig/ele4-biaxialtension.pdf_tex
\begingroup%
  \makeatletter%
  \providecommand\color[2][]{%
    \errmessage{(Inkscape) Color is used for the text in Inkscape, but the package 'color.sty' is not loaded}%
    \renewcommand\color[2][]{}%
  }%
  \providecommand\transparent[1]{%
    \errmessage{(Inkscape) Transparency is used (non-zero) for the text in Inkscape, but the package 'transparent.sty' is not loaded}%
    \renewcommand\transparent[1]{}%
  }%
  \providecommand\rotatebox[2]{#2}%
  \ifx\svgwidth\undefined%
    \setlength{\unitlength}{174.38816117bp}%
    \ifx\svgscale\undefined%
      \relax%
    \else%
      \setlength{\unitlength}{\unitlength * \real{\svgscale}}%
    \fi%
  \else%
    \setlength{\unitlength}{\svgwidth}%
  \fi%
  \global\let\svgwidth\undefined%
  \global\let\svgscale\undefined%
  \makeatother%
  \begin{picture}(1,0.82364282)%
    \put(0,0){\includegraphics[width=\unitlength]{ele4-biaxialtension.pdf}}%
    \put(0.03442604,0.75393482){\color[rgb]{0,0,0}\makebox(0,0)[lb]{\smash{$y$}}}%
    \put(0.83500349,0.04160998){\color[rgb]{0,0,0}\makebox(0,0)[lb]{\smash{$x$}}}%
    \put(0.60648341,0.03997423){\color[rgb]{0,0,0}\makebox(0,0)[lb]{\smash{$u_x$}}}%
    \put(0.01811916,0.57608828){\color[rgb]{0,0,0}\makebox(0,0)[lb]{\smash{$u_y$}}}%
    \put(0.5450307,0.54086791){\color[rgb]{0,0,0}\makebox(0,0)[lb]{\smash{$A$}}}%
    \put(0.73288264,0.70728708){\color[rgb]{0,0,0}\makebox(0,0)[lb]{\smash{$A'$}}}%
  \end{picture}%
\endgroup%

%% file: Fig/circle.pdf_tex
\begingroup%
  \makeatletter%
  \providecommand\color[2][]{%
    \errmessage{(Inkscape) Color is used for the text in Inkscape, but the package 'color.sty' is not loaded}%
    \renewcommand\color[2][]{}%
  }%
  \providecommand\transparent[1]{%
    \errmessage{(Inkscape) Transparency is used (non-zero) for the text in Inkscape, but the package 'transparent.sty' is not loaded}%
    \renewcommand\transparent[1]{}%
  }%
  \providecommand\rotatebox[2]{#2}%
  \ifx\svgwidth\undefined%
    \setlength{\unitlength}{227.4902832bp}%
    \ifx\svgscale\undefined%
      \relax%
    \else%
      \setlength{\unitlength}{\unitlength * \real{\svgscale}}%
    \fi%
  \else%
    \setlength{\unitlength}{\svgwidth}%
  \fi%
  \global\let\svgwidth\undefined%
  \global\let\svgscale\undefined%
  \makeatother%
  \begin{picture}(1,0.8667969)%
    \put(0,0){\includegraphics[width=\unitlength]{circle.pdf}}%
    \put(0.48860711,0.83188338){\color[rgb]{0,0,0}\makebox(0,0)[lb]{\smash{$u_y$}}}%
    \put(0.89685342,0.4421772){\color[rgb]{0,0,0}\makebox(0,0)[lb]{\smash{$u_x$}}}%
    \put(0.54269896,0.42864683){\color[rgb]{0,0,0}\makebox(0,0)[lb]{\smash{$\phi$}}}%
    \put(0.40029676,0.41465338){\color[rgb]{0,0,0}\makebox(0,0)[lb]{\smash{$O$}}}%
    \put(0.73086323,0.64001403){\color[rgb]{0,0,0}\makebox(0,0)[lb]{\smash{$P_i$}}}%
    \put(0.32450068,0.66588181){\color[rgb]{0,0,0}\makebox(0,0)[lb]{\smash{$P_j$}}}%
    \put(0.3254672,0.22082717){\color[rgb]{0,0,0}\makebox(0,0)[lb]{\smash{$P_k$}}}%
    \put(0.68575677,0.68763958){\color[rgb]{0,0,0}\makebox(0,0)[lb]{\smash{Loading-unloading path $OP_i$}}}%
    \put(0.80921725,0.41653522){\color[rgb]{0,0,0}\makebox(0,0)[lb]{\smash{$r_1$}}}%
    \put(0.7183159,0.41750163){\color[rgb]{0,0,0}\makebox(0,0)[lb]{\smash{$r_2$}}}%
    \put(0.63133449,0.41846804){\color[rgb]{0,0,0}\makebox(0,0)[lb]{\smash{$r_3$}}}%
  \end{picture}%
\endgroup%

%% file: Fig/ele8-biaxialtension.pdf_tex
\begingroup%
  \makeatletter%
  \providecommand\color[2][]{%
    \errmessage{(Inkscape) Color is used for the text in Inkscape, but the package 'color.sty' is not loaded}%
    \renewcommand\color[2][]{}%
  }%
  \providecommand\transparent[1]{%
    \errmessage{(Inkscape) Transparency is used (non-zero) for the text in Inkscape, but the package 'transparent.sty' is not loaded}%
    \renewcommand\transparent[1]{}%
  }%
  \providecommand\rotatebox[2]{#2}%
  \ifx\svgwidth\undefined%
    \setlength{\unitlength}{249.88945317bp}%
    \ifx\svgscale\undefined%
      \relax%
    \else%
      \setlength{\unitlength}{\unitlength * \real{\svgscale}}%
    \fi%
  \else%
    \setlength{\unitlength}{\svgwidth}%
  \fi%
  \global\let\svgwidth\undefined%
  \global\let\svgscale\undefined%
  \makeatother%
  \begin{picture}(1,0.95403474)%
    \put(0,0){\includegraphics[width=\unitlength]{ele8-biaxialtension.pdf}}%
    \put(0.30831651,0.90467992){\color[rgb]{0,0,0}\makebox(0,0)[lb]{\smash{$z$}}}%
    \put(0.03708212,0.00941041){\color[rgb]{0,0,0}\makebox(0,0)[lb]{\smash{$x$}}}%
    \put(0.14512433,0.1501154){\color[rgb]{0,0,0}\makebox(0,0)[lb]{\smash{$u_x$}}}%
    \put(0.29320377,0.75319379){\color[rgb]{0,0,0}\makebox(0,0)[lb]{\smash{$u_z$}}}%
    \put(0.44921289,0.4755258){\color[rgb]{0,0,0}\makebox(0,0)[lb]{\smash{$A$}}}%
    \put(0.55684291,0.53555185){\color[rgb]{0,0,0}\makebox(0,0)[lb]{\smash{$A'$}}}%
    \put(0.88485533,0.39062927){\color[rgb]{0,0,0}\makebox(0,0)[lb]{\smash{$y$}}}%
    \put(0.69174914,0.38476428){\color[rgb]{0,0,0}\makebox(0,0)[lb]{\smash{$u_y$}}}%
  \end{picture}%
\endgroup%

%% file: Fig/sphere.pdf_tex
\begingroup%
  \makeatletter%
  \providecommand\color[2][]{%
    \errmessage{(Inkscape) Color is used for the text in Inkscape, but the package 'color.sty' is not loaded}%
    \renewcommand\color[2][]{}%
  }%
  \providecommand\transparent[1]{%
    \errmessage{(Inkscape) Transparency is used (non-zero) for the text in Inkscape, but the package 'transparent.sty' is not loaded}%
    \renewcommand\transparent[1]{}%
  }%
  \providecommand\rotatebox[2]{#2}%
  \ifx\svgwidth\undefined%
    \setlength{\unitlength}{295.50985615bp}%
    \ifx\svgscale\undefined%
      \relax%
    \else%
      \setlength{\unitlength}{\unitlength * \real{\svgscale}}%
    \fi%
  \else%
    \setlength{\unitlength}{\svgwidth}%
  \fi%
  \global\let\svgwidth\undefined%
  \global\let\svgscale\undefined%
  \makeatother%
  \begin{picture}(1,0.98150876)%
    \put(0,0){\includegraphics[width=\unitlength]{sphere.pdf}}%
    \put(0.05338882,0.01276924){\color[rgb]{0,0,0}\makebox(0,0)[lb]{\smash{$u_x$}}}%
    \put(0.3548427,0.44365622){\color[rgb]{0,0,0}\makebox(0,0)[lb]{\smash{$O$}}}%
    \put(0.84119079,0.32005958){\color[rgb]{0,0,0}\makebox(0,0)[lb]{\smash{$u_y$}}}%
    \put(0.41611028,0.93242787){\color[rgb]{0,0,0}\makebox(0,0)[lb]{\smash{$u_z$}}}%
    \put(0.53000152,0.42037449){\color[rgb]{0,0,0}\makebox(0,0)[lb]{\smash{$\theta$}}}%
    \put(0.59734044,0.52891761){\color[rgb]{0,0,0}\makebox(0,0)[lb]{\smash{$P_i$}}}%
    \put(0.51502103,0.60162852){\color[rgb]{0,0,0}\makebox(0,0)[lb]{\smash{Loading-unloading path $OP_i$}}}%
    \put(0.38375096,0.34452009){\color[rgb]{0,0,0}\makebox(0,0)[lb]{\smash{$\phi$}}}%
  \end{picture}%
\endgroup%

%% file: Fig/FNN.pdf_tex
\begingroup%
  \makeatletter%
  \providecommand\color[2][]{%
    \errmessage{(Inkscape) Color is used for the text in Inkscape, but the package 'color.sty' is not loaded}%
    \renewcommand\color[2][]{}%
  }%
  \providecommand\transparent[1]{%
    \errmessage{(Inkscape) Transparency is used (non-zero) for the text in Inkscape, but the package 'transparent.sty' is not loaded}%
    \renewcommand\transparent[1]{}%
  }%
  \providecommand\rotatebox[2]{#2}%
  \ifx\svgwidth\undefined%
    \setlength{\unitlength}{237.18249512bp}%
    \ifx\svgscale\undefined%
      \relax%
    \else%
      \setlength{\unitlength}{\unitlength * \real{\svgscale}}%
    \fi%
  \else%
    \setlength{\unitlength}{\svgwidth}%
  \fi%
  \global\let\svgwidth\undefined%
  \global\let\svgscale\undefined%
  \makeatother%
  \begin{picture}(1,0.25486601)%
    \put(0,0){\includegraphics[width=\unitlength]{FNN.pdf}}%
    \put(0.39445831,0.14153872){\color[rgb]{0,0,0}\makebox(0,0)[lb]{\smash{$\sigma_1^t$}}}%
    \put(0.39436171,0.11063392){\color[rgb]{0,0,0}\makebox(0,0)[lb]{\smash{$\sigma_2^t$}}}%
    \put(0.39395203,0.07968156){\color[rgb]{0,0,0}\makebox(0,0)[lb]{\smash{$\sigma_3^t$}}}%
    \put(0.21804489,0.24464264){\color[rgb]{0,0,0}\makebox(0,0)[lb]{\smash{FNN}}}%
    \put(0.55984084,0.09104832){\color[rgb]{0,0,0}\makebox(0,0)[lb]{\smash{$\varepsilon_{acce,1}^t$}}}%
    \put(0.95732782,0.11327685){\color[rgb]{0,0,0}\makebox(0,0)[lb]{\smash{$\alpha_i^t$}}}%
    \put(0.5607902,0.05219622){\color[rgb]{0,0,0}\makebox(0,0)[lb]{\smash{$\varepsilon_{acce,2}^t$}}}%
    \put(0.56171712,0.01816841){\color[rgb]{0,0,0}\makebox(0,0)[lb]{\smash{$\varepsilon_{acce,3}^t$}}}%
    \put(0.56024723,0.19915749){\color[rgb]{0,0,0}\makebox(0,0)[lb]{\smash{$\varepsilon_1^t$}}}%
    \put(0.55971063,0.16696107){\color[rgb]{0,0,0}\makebox(0,0)[lb]{\smash{$\varepsilon_2^t$}}}%
    \put(0.56033801,0.1308083){\color[rgb]{0,0,0}\makebox(0,0)[lb]{\smash{$\varepsilon_3^t$}}}%
    \put(0.73716476,0.24974074){\color[rgb]{0,0,0}\makebox(0,0)[lb]{\smash{FNNi   $(i=1,2,3.)$  }}}%
    \put(-0.00009377,0.0929598){\color[rgb]{0,0,0}\makebox(0,0)[lb]{\smash{$\varepsilon_{acce,1}^t$}}}%
    \put(0.00085558,0.0541077){\color[rgb]{0,0,0}\makebox(0,0)[lb]{\smash{$\varepsilon_{acce,2}^t$}}}%
    \put(0.0017825,0.02007989){\color[rgb]{0,0,0}\makebox(0,0)[lb]{\smash{$\varepsilon_{acce,3}^t$}}}%
    \put(0.00031261,0.20106897){\color[rgb]{0,0,0}\makebox(0,0)[lb]{\smash{$\varepsilon_1^t$}}}%
    \put(-0.00022398,0.16887255){\color[rgb]{0,0,0}\makebox(0,0)[lb]{\smash{$\varepsilon_2^t$}}}%
    \put(0.0004034,0.13271978){\color[rgb]{0,0,0}\makebox(0,0)[lb]{\smash{$\varepsilon_3^t$}}}%
  \end{picture}%
\endgroup%

%% file: Fig/POD1D.pdf_tex
\begingroup%
  \makeatletter%
  \providecommand\color[2][]{%
    \errmessage{(Inkscape) Color is used for the text in Inkscape, but the package 'color.sty' is not loaded}%
    \renewcommand\color[2][]{}%
  }%
  \providecommand\transparent[1]{%
    \errmessage{(Inkscape) Transparency is used (non-zero) for the text in Inkscape, but the package 'transparent.sty' is not loaded}%
    \renewcommand\transparent[1]{}%
  }%
  \providecommand\rotatebox[2]{#2}%
  \ifx\svgwidth\undefined%
    \setlength{\unitlength}{480bp}%
    \ifx\svgscale\undefined%
      \relax%
    \else%
      \setlength{\unitlength}{\unitlength * \real{\svgscale}}%
    \fi%
  \else%
    \setlength{\unitlength}{\svgwidth}%
  \fi%
  \global\let\svgwidth\undefined%
  \global\let\svgscale\undefined%
  \makeatother%
  \begin{picture}(1,0.8)%
    \put(0,0){\includegraphics[width=\unitlength]{POD1D.pdf}}%
    \put(0.106,0.16466667){\makebox(0,0)[rb]{\smash{-150}}}%
    \put(0.106,0.2585){\makebox(0,0)[rb]{\smash{-100}}}%
    \put(0.106,0.3525){\makebox(0,0)[rb]{\smash{-50}}}%
    \put(0.106,0.44633333){\makebox(0,0)[rb]{\smash{0}}}%
    \put(0.106,0.54033333){\makebox(0,0)[rb]{\smash{50}}}%
    \put(0.106,0.63416667){\makebox(0,0)[rb]{\smash{100}}}%
    \put(0.106,0.72816667){\makebox(0,0)[rb]{\smash{150}}}%
    \put(0.11983333,0.0595){\makebox(0,0)[b]{\smash{-1.2}}}%
    \put(0.2595,0.0595){\makebox(0,0)[b]{\smash{-0.8}}}%
    \put(0.39933333,0.0595){\makebox(0,0)[b]{\smash{-0.4}}}%
    \put(0.539,0.0595){\makebox(0,0)[b]{\smash{0}}}%
    \put(0.67883333,0.0595){\makebox(0,0)[b]{\smash{0.4}}}%
    \put(0.81866667,0.0595){\makebox(0,0)[b]{\smash{0.8}}}%
    \put(0.95833333,0.0595){\makebox(0,0)[b]{\smash{1.2}}}%
    \put(0.02833333,0.434){\rotatebox{90}{\makebox(0,0)[b]{\smash{Stress}}}}%
    \put(0.539,0.0145){\makebox(0,0)[b]{\smash{Strain}}}%
    \put(0.7895,0.17516667){\makebox(0,0)[rb]{\smash{Plasticity}}}%
    \put(0.7895,0.12516667){\makebox(0,0)[rb]{\smash{NN model}}}%
  \end{picture}%
\endgroup%

%% file: Fig/cook2D-uf-unloading.pdf_tex
\begingroup%
  \makeatletter%
  \providecommand\color[2][]{%
    \errmessage{(Inkscape) Color is used for the text in Inkscape, but the package 'color.sty' is not loaded}%
    \renewcommand\color[2][]{}%
  }%
  \providecommand\transparent[1]{%
    \errmessage{(Inkscape) Transparency is used (non-zero) for the text in Inkscape, but the package 'transparent.sty' is not loaded}%
    \renewcommand\transparent[1]{}%
  }%
  \providecommand\rotatebox[2]{#2}%
  \ifx\svgwidth\undefined%
    \setlength{\unitlength}{480bp}%
    \ifx\svgscale\undefined%
      \relax%
    \else%
      \setlength{\unitlength}{\unitlength * \real{\svgscale}}%
    \fi%
  \else%
    \setlength{\unitlength}{\svgwidth}%
  \fi%
  \global\let\svgwidth\undefined%
  \global\let\svgscale\undefined%
  \makeatother%
  \begin{picture}(1,0.8)%
    \put(0,0){\includegraphics[width=\unitlength]{cook2D-uf-unloading.pdf}}%
    \put(0.106,0.1195){\makebox(0,0)[rb]{\smash{0}}}%
    \put(0.106,0.281){\makebox(0,0)[rb]{\smash{0.01}}}%
    \put(0.106,0.44266667){\makebox(0,0)[rb]{\smash{0.02}}}%
    \put(0.106,0.60416667){\makebox(0,0)[rb]{\smash{0.03}}}%
    \put(0.106,0.76566667){\makebox(0,0)[rb]{\smash{0.04}}}%
    \put(0.11983333,0.0895){\makebox(0,0)[b]{\smash{0}}}%
    \put(0.25966667,0.0895){\makebox(0,0)[b]{\smash{2}}}%
    \put(0.39933333,0.0895){\makebox(0,0)[b]{\smash{4}}}%
    \put(0.53916667,0.0895){\makebox(0,0)[b]{\smash{6}}}%
    \put(0.67883333,0.0895){\makebox(0,0)[b]{\smash{8}}}%
    \put(0.81866667,0.0895){\makebox(0,0)[b]{\smash{10}}}%
    \put(0.95833333,0.0895){\makebox(0,0)[b]{\smash{12}}}%
    \put(0.02833333,0.449){\rotatebox{90}{\makebox(0,0)[b]{\smash{Force $Fy$}}}}%
    \put(0.539,0.0445){\makebox(0,0)[b]{\smash{Displacement $Uy$}}}%
    \put(0.7895,0.20516667){\makebox(0,0)[rb]{\smash{Plasticity}}}%
    \put(0.7895,0.15516667){\makebox(0,0)[rb]{\smash{NN model}}}%
  \end{picture}%
\endgroup%

%% file: Fig/punch.pdf_tex
\begingroup%
  \makeatletter%
  \providecommand\color[2][]{%
    \errmessage{(Inkscape) Color is used for the text in Inkscape, but the package 'color.sty' is not loaded}%
    \renewcommand\color[2][]{}%
  }%
  \providecommand\transparent[1]{%
    \errmessage{(Inkscape) Transparency is used (non-zero) for the text in Inkscape, but the package 'transparent.sty' is not loaded}%
    \renewcommand\transparent[1]{}%
  }%
  \providecommand\rotatebox[2]{#2}%
  \ifx\svgwidth\undefined%
    \setlength{\unitlength}{228.13659668bp}%
    \ifx\svgscale\undefined%
      \relax%
    \else%
      \setlength{\unitlength}{\unitlength * \real{\svgscale}}%
    \fi%
  \else%
    \setlength{\unitlength}{\svgwidth}%
  \fi%
  \global\let\svgwidth\undefined%
  \global\let\svgscale\undefined%
  \makeatother%
  \begin{picture}(1,0.99750141)%
    \put(0,0){\includegraphics[width=\unitlength]{punch.pdf}}%
    \put(0.30319124,0.0103077){\color[rgb]{0,0,0}\makebox(0,0)[lb]{\smash{$1mm$}}}%
    \put(0.79114075,0.63212315){\color[rgb]{0,0,0}\makebox(0,0)[lb]{\smash{$u_0$}}}%
    \put(0.72954607,0.47824891){\color[rgb]{0,0,0}\makebox(0,0)[lb]{\smash{$1mm$}}}%
    \put(0.84532711,0.19325387){\color[rgb]{0,0,0}\makebox(0,0)[lb]{\smash{$x$}}}%
    \put(0.06595582,0.94421645){\color[rgb]{0,0,0}\makebox(0,0)[lb]{\smash{$y$}}}%
  \end{picture}%
\endgroup%

%% file: Fig/punch2D-uf-unloading.pdf_tex
\begingroup%
  \makeatletter%
  \providecommand\color[2][]{%
    \errmessage{(Inkscape) Color is used for the text in Inkscape, but the package 'color.sty' is not loaded}%
    \renewcommand\color[2][]{}%
  }%
  \providecommand\transparent[1]{%
    \errmessage{(Inkscape) Transparency is used (non-zero) for the text in Inkscape, but the package 'transparent.sty' is not loaded}%
    \renewcommand\transparent[1]{}%
  }%
  \providecommand\rotatebox[2]{#2}%
  \ifx\svgwidth\undefined%
    \setlength{\unitlength}{480bp}%
    \ifx\svgscale\undefined%
      \relax%
    \else%
      \setlength{\unitlength}{\unitlength * \real{\svgscale}}%
    \fi%
  \else%
    \setlength{\unitlength}{\svgwidth}%
  \fi%
  \global\let\svgwidth\undefined%
  \global\let\svgscale\undefined%
  \makeatother%
  \begin{picture}(1,0.8)%
    \put(0,0){\includegraphics[width=\unitlength]{punch2D-uf-unloading.pdf}}%
    \put(0.13366667,0.1195){\makebox(0,0)[rb]{\smash{0}}}%
    \put(0.13366667,0.22716667){\makebox(0,0)[rb]{\smash{0.0002}}}%
    \put(0.13366667,0.33483333){\makebox(0,0)[rb]{\smash{0.0004}}}%
    \put(0.13366667,0.44266667){\makebox(0,0)[rb]{\smash{0.0006}}}%
    \put(0.13366667,0.55033333){\makebox(0,0)[rb]{\smash{0.0008}}}%
    \put(0.13366667,0.658){\makebox(0,0)[rb]{\smash{0.001}}}%
    \put(0.13366667,0.76566667){\makebox(0,0)[rb]{\smash{0.0012}}}%
    \put(0.1475,0.0895){\makebox(0,0)[b]{\smash{0}}}%
    \put(0.35016667,0.0895){\makebox(0,0)[b]{\smash{0.02}}}%
    \put(0.553,0.0895){\makebox(0,0)[b]{\smash{0.04}}}%
    \put(0.75566667,0.0895){\makebox(0,0)[b]{\smash{0.06}}}%
    \put(0.95833333,0.0895){\makebox(0,0)[b]{\smash{0.08}}}%
    \put(0.02833333,0.449){\rotatebox{90}{\makebox(0,0)[b]{\smash{Force $Fy$}}}}%
    \put(0.55283333,0.0445){\makebox(0,0)[b]{\smash{Displacement $Uy$}}}%
    \put(0.7895,0.20516667){\makebox(0,0)[rb]{\smash{Plasticity}}}%
    \put(0.7895,0.15516667){\makebox(0,0)[rb]{\smash{NN model}}}%
  \end{picture}%
\endgroup%

%% file: Fig/bar3Dgeo.pdf_tex
\begingroup%
  \makeatletter%
  \providecommand\color[2][]{%
    \errmessage{(Inkscape) Color is used for the text in Inkscape, but the package 'color.sty' is not loaded}%
    \renewcommand\color[2][]{}%
  }%
  \providecommand\transparent[1]{%
    \errmessage{(Inkscape) Transparency is used (non-zero) for the text in Inkscape, but the package 'transparent.sty' is not loaded}%
    \renewcommand\transparent[1]{}%
  }%
  \providecommand\rotatebox[2]{#2}%
  \ifx\svgwidth\undefined%
    \setlength{\unitlength}{371.60064745bp}%
    \ifx\svgscale\undefined%
      \relax%
    \else%
      \setlength{\unitlength}{\unitlength * \real{\svgscale}}%
    \fi%
  \else%
    \setlength{\unitlength}{\svgwidth}%
  \fi%
  \global\let\svgwidth\undefined%
  \global\let\svgscale\undefined%
  \makeatother%
  \begin{picture}(1,0.33271053)%
    \put(0,0){\includegraphics[width=\unitlength]{bar3Dgeo.pdf}}%
    \put(0.36715404,0.0063282){\color[rgb]{0,0,0}\makebox(0,0)[lb]{\smash{$L=10mm$}}}%
    \put(0.06632904,0.20018655){\color[rgb]{0,0,0}\makebox(0,0)[lb]{\smash{$R=0.5$}}}%
    \put(0.42032083,0.29999732){\color[rgb]{0,0,0}\makebox(0,0)[lb]{\smash{$y$}}}%
    \put(0.49341139,0.18309267){\color[rgb]{0,0,0}\makebox(0,0)[lb]{\smash{$z$}}}%
    \put(0.31327067,0.11484037){\color[rgb]{0,0,0}\makebox(0,0)[lb]{\smash{$x$}}}%
    \put(0.76421681,0.27824217){\color[rgb]{0,0,0}\makebox(0,0)[lb]{\smash{$u_z=u_0$}}}%
    \put(0.00005411,0.27573192){\color[rgb]{0,0,0}\makebox(0,0)[lb]{\smash{$u_z=0$}}}%
  \end{picture}%
\endgroup%

%% file: Fig/bar3D-uf.pdf_tex
\begingroup%
  \makeatletter%
  \providecommand\color[2][]{%
    \errmessage{(Inkscape) Color is used for the text in Inkscape, but the package 'color.sty' is not loaded}%
    \renewcommand\color[2][]{}%
  }%
  \providecommand\transparent[1]{%
    \errmessage{(Inkscape) Transparency is used (non-zero) for the text in Inkscape, but the package 'transparent.sty' is not loaded}%
    \renewcommand\transparent[1]{}%
  }%
  \providecommand\rotatebox[2]{#2}%
  \ifx\svgwidth\undefined%
    \setlength{\unitlength}{480bp}%
    \ifx\svgscale\undefined%
      \relax%
    \else%
      \setlength{\unitlength}{\unitlength * \real{\svgscale}}%
    \fi%
  \else%
    \setlength{\unitlength}{\svgwidth}%
  \fi%
  \global\let\svgwidth\undefined%
  \global\let\svgscale\undefined%
  \makeatother%
  \begin{picture}(1,0.8)%
    \put(0,0){\includegraphics[width=\unitlength]{bar3D-uf.pdf}}%
    \put(0.13366667,0.1195){\makebox(0,0)[rb]{\smash{0.0e0}}}%
    \put(0.13366667,0.281){\makebox(0,0)[rb]{\smash{1.0e-4}}}%
    \put(0.13366667,0.44266667){\makebox(0,0)[rb]{\smash{2.0e-4}}}%
    \put(0.13366667,0.60416667){\makebox(0,0)[rb]{\smash{3.0e-4}}}%
    \put(0.13366667,0.76566667){\makebox(0,0)[rb]{\smash{4.0e-4}}}%
    \put(0.1475,0.0895){\makebox(0,0)[b]{\smash{0}}}%
    \put(0.30966667,0.0895){\makebox(0,0)[b]{\smash{0.01}}}%
    \put(0.47183333,0.0895){\makebox(0,0)[b]{\smash{0.02}}}%
    \put(0.634,0.0895){\makebox(0,0)[b]{\smash{0.03}}}%
    \put(0.79616667,0.0895){\makebox(0,0)[b]{\smash{0.04}}}%
    \put(0.95833333,0.0895){\makebox(0,0)[b]{\smash{0.05}}}%
    \put(0.02114242,0.449){\rotatebox{90}{\makebox(0,0)[b]{\smash{Force $F_z$}}}}%
    \put(0.55283333,0.0445){\makebox(0,0)[b]{\smash{Displacement $U_z$}}}%
    \put(0.7895,0.20516667){\makebox(0,0)[rb]{\smash{Plasticity}}}%
    \put(0.7895,0.15516667){\makebox(0,0)[rb]{\smash{NN model}}}%
  \end{picture}%
\endgroup%

%% file: Fig/punch3Dgeo.pdf_tex
\begingroup%
  \makeatletter%
  \providecommand\color[2][]{%
    \errmessage{(Inkscape) Color is used for the text in Inkscape, but the package 'color.sty' is not loaded}%
    \renewcommand\color[2][]{}%
  }%
  \providecommand\transparent[1]{%
    \errmessage{(Inkscape) Transparency is used (non-zero) for the text in Inkscape, but the package 'transparent.sty' is not loaded}%
    \renewcommand\transparent[1]{}%
  }%
  \providecommand\rotatebox[2]{#2}%
  \ifx\svgwidth\undefined%
    \setlength{\unitlength}{282.45046387bp}%
    \ifx\svgscale\undefined%
      \relax%
    \else%
      \setlength{\unitlength}{\unitlength * \real{\svgscale}}%
    \fi%
  \else%
    \setlength{\unitlength}{\svgwidth}%
  \fi%
  \global\let\svgwidth\undefined%
  \global\let\svgscale\undefined%
  \makeatother%
  \begin{picture}(1,0.90995803)%
    \put(0,0){\includegraphics[width=\unitlength]{punch3Dgeo.pdf}}%
    \put(0.23614717,0.00832557){\color[rgb]{0,0,0}\makebox(0,0)[lb]{\smash{$1mm$}}}%
    \put(0.34243664,0.86447499){\color[rgb]{0,0,0}\makebox(0,0)[lb]{\smash{$u_0$}}}%
    \put(0.71005079,0.1351803){\color[rgb]{0,0,0}\makebox(0,0)[lb]{\smash{$1mm$}}}%
    \put(0.78516698,0.52519174){\color[rgb]{0,0,0}\makebox(0,0)[lb]{\smash{$1mm$}}}%
    \put(0.61623672,0.15421261){\color[rgb]{0,0,0}\makebox(0,0)[lb]{\smash{$x$}}}%
    \put(0.04798343,0.81180311){\color[rgb]{0,0,0}\makebox(0,0)[lb]{\smash{$z$}}}%
  \end{picture}%
\endgroup%

%% file: Fig/punch3D-uf.pdf_tex
\begingroup%
  \makeatletter%
  \providecommand\color[2][]{%
    \errmessage{(Inkscape) Color is used for the text in Inkscape, but the package 'color.sty' is not loaded}%
    \renewcommand\color[2][]{}%
  }%
  \providecommand\transparent[1]{%
    \errmessage{(Inkscape) Transparency is used (non-zero) for the text in Inkscape, but the package 'transparent.sty' is not loaded}%
    \renewcommand\transparent[1]{}%
  }%
  \providecommand\rotatebox[2]{#2}%
  \ifx\svgwidth\undefined%
    \setlength{\unitlength}{480bp}%
    \ifx\svgscale\undefined%
      \relax%
    \else%
      \setlength{\unitlength}{\unitlength * \real{\svgscale}}%
    \fi%
  \else%
    \setlength{\unitlength}{\svgwidth}%
  \fi%
  \global\let\svgwidth\undefined%
  \global\let\svgscale\undefined%
  \makeatother%
  \begin{picture}(1,0.8)%
    \put(0,0){\includegraphics[width=\unitlength]{punch3D-uf.pdf}}%
    \put(0.13366667,0.1195){\makebox(0,0)[rb]{\smash{0.0e0}}}%
    \put(0.13366667,0.22716667){\makebox(0,0)[rb]{\smash{2.0e-4}}}%
    \put(0.13366667,0.33483333){\makebox(0,0)[rb]{\smash{4.0e-4}}}%
    \put(0.13366667,0.44266667){\makebox(0,0)[rb]{\smash{6.0e-4}}}%
    \put(0.13366667,0.55033333){\makebox(0,0)[rb]{\smash{8.0e-4}}}%
    \put(0.13366667,0.658){\makebox(0,0)[rb]{\smash{1.0e-3}}}%
    \put(0.13366667,0.76566667){\makebox(0,0)[rb]{\smash{1.2e-3}}}%
    \put(0.1475,0.0895){\makebox(0,0)[b]{\smash{0}}}%
    \put(0.28266667,0.0895){\makebox(0,0)[b]{\smash{0.02}}}%
    \put(0.41783333,0.0895){\makebox(0,0)[b]{\smash{0.04}}}%
    \put(0.553,0.0895){\makebox(0,0)[b]{\smash{0.06}}}%
    \put(0.688,0.0895){\makebox(0,0)[b]{\smash{0.08}}}%
    \put(0.82316667,0.0895){\makebox(0,0)[b]{\smash{0.1}}}%
    \put(0.95833333,0.0895){\makebox(0,0)[b]{\smash{0.12}}}%
    \put(0.02833333,0.449){\rotatebox{90}{\makebox(0,0)[b]{\smash{Force $F_z$}}}}%
    \put(0.55283333,0.0445){\makebox(0,0)[b]{\smash{Displacement $U_z$}}}%
    \put(0.7895,0.20516667){\makebox(0,0)[rb]{\smash{Plasticity}}}%
    \put(0.7895,0.15516667){\makebox(0,0)[rb]{\smash{NN model}}}%
  \end{picture}%
\endgroup%

%% file: Fig/cook3Dgeo.pdf_tex
\begingroup%
  \makeatletter%
  \providecommand\color[2][]{%
    \errmessage{(Inkscape) Color is used for the text in Inkscape, but the package 'color.sty' is not loaded}%
    \renewcommand\color[2][]{}%
  }%
  \providecommand\transparent[1]{%
    \errmessage{(Inkscape) Transparency is used (non-zero) for the text in Inkscape, but the package 'transparent.sty' is not loaded}%
    \renewcommand\transparent[1]{}%
  }%
  \providecommand\rotatebox[2]{#2}%
  \ifx\svgwidth\undefined%
    \setlength{\unitlength}{290.82171631bp}%
    \ifx\svgscale\undefined%
      \relax%
    \else%
      \setlength{\unitlength}{\unitlength * \real{\svgscale}}%
    \fi%
  \else%
    \setlength{\unitlength}{\svgwidth}%
  \fi%
  \global\let\svgwidth\undefined%
  \global\let\svgscale\undefined%
  \makeatother%
  \begin{picture}(1,0.68334593)%
    \put(0,0){\includegraphics[width=\unitlength]{cook3Dgeo.pdf}}%
    \put(0.49303393,0.11865238){\color[rgb]{0,0,0}\makebox(0,0)[lb]{\smash{$x$}}}%
    \put(0.22163107,0.5835902){\color[rgb]{0,0,0}\makebox(0,0)[lb]{\smash{$z$}}}%
    \put(0.44057909,0.46255015){\color[rgb]{0,0,0}\makebox(0,0)[lb]{\smash{$48mm$}}}%
    \put(0.28103143,0.23119667){\color[rgb]{0,0,0}\makebox(0,0)[lb]{\smash{$44mm$}}}%
    \put(0.75632085,0.52092747){\color[rgb]{0,0,0}\makebox(0,0)[lb]{\smash{$16mm$}}}%
    \put(0.69924742,0.64154627){\color[rgb]{0,0,0}\makebox(0,0)[lb]{\smash{$q_0$}}}%
    \put(0.42181223,0.55152429){\color[rgb]{0,0,0}\makebox(0,0)[lb]{\smash{$22mm$}}}%
    \put(0.03063312,0.12447522){\color[rgb]{0,0,0}\makebox(0,0)[lb]{\smash{$y$}}}%
    \put(-0.00091336,0.27113855){\color[rgb]{0,0,0}\makebox(0,0)[lb]{\smash{$u_z=0$}}}%
    \put(0.02342819,0.40815031){\color[rgb]{0,0,0}\makebox(0,0)[lb]{\smash{$u_y=0$}}}%
  \end{picture}%
\endgroup%

%% file: Fig/cook3D-uf.pdf_tex
\begingroup%
  \makeatletter%
  \providecommand\color[2][]{%
    \errmessage{(Inkscape) Color is used for the text in Inkscape, but the package 'color.sty' is not loaded}%
    \renewcommand\color[2][]{}%
  }%
  \providecommand\transparent[1]{%
    \errmessage{(Inkscape) Transparency is used (non-zero) for the text in Inkscape, but the package 'transparent.sty' is not loaded}%
    \renewcommand\transparent[1]{}%
  }%
  \providecommand\rotatebox[2]{#2}%
  \ifx\svgwidth\undefined%
    \setlength{\unitlength}{480bp}%
    \ifx\svgscale\undefined%
      \relax%
    \else%
      \setlength{\unitlength}{\unitlength * \real{\svgscale}}%
    \fi%
  \else%
    \setlength{\unitlength}{\svgwidth}%
  \fi%
  \global\let\svgwidth\undefined%
  \global\let\svgscale\undefined%
  \makeatother%
  \begin{picture}(1,0.8)%
    \put(0,0){\includegraphics[width=\unitlength]{cook3D-uf.pdf}}%
    \put(0.106,0.1195){\makebox(0,0)[rb]{\smash{0}}}%
    \put(0.106,0.26316667){\makebox(0,0)[rb]{\smash{0.04}}}%
    \put(0.106,0.40666667){\makebox(0,0)[rb]{\smash{0.08}}}%
    \put(0.106,0.55033333){\makebox(0,0)[rb]{\smash{0.12}}}%
    \put(0.106,0.69383333){\makebox(0,0)[rb]{\smash{0.16}}}%
    \put(0.11983333,0.0895){\makebox(0,0)[b]{\smash{0}}}%
    \put(0.25966667,0.0895){\makebox(0,0)[b]{\smash{2}}}%
    \put(0.39933333,0.0895){\makebox(0,0)[b]{\smash{4}}}%
    \put(0.53916667,0.0895){\makebox(0,0)[b]{\smash{6}}}%
    \put(0.67883333,0.0895){\makebox(0,0)[b]{\smash{8}}}%
    \put(0.81866667,0.0895){\makebox(0,0)[b]{\smash{10}}}%
    \put(0.95833333,0.0895){\makebox(0,0)[b]{\smash{12}}}%
    \put(0.02833333,0.449){\rotatebox{90}{\makebox(0,0)[b]{\smash{Force $F_z$}}}}%
    \put(0.539,0.0445){\makebox(0,0)[b]{\smash{Displacement $U_z$}}}%
    \put(0.7895,0.20516667){\makebox(0,0)[rb]{\smash{Plasticity}}}%
    \put(0.7895,0.15516667){\makebox(0,0)[rb]{\smash{NN model}}}%
  \end{picture}%
\endgroup%